\documentclass[notitlepage]{revtex4-1}
\usepackage{setspace}
\usepackage[utf8]{inputenc}
\usepackage[dvipsnames]{xcolor}
\usepackage[colorlinks=true,citecolor=RedViolet,urlcolor=BlueGreen,linkcolor=RedViolet]{hyperref}
\usepackage[normalem]{ulem}
\usepackage{url}
\usepackage{mathrsfs}
\usepackage{amsmath}
\usepackage{graphicx,wrapfig,float,slashed,cancel}
\usepackage{amsmath,amssymb,epsfig,xcolor,stmaryrd}
\usepackage{bm}
\usepackage{enumitem}
\usepackage{hhline,multirow,tabularx} 
\usepackage{graphics}
\usepackage{latexsym}
\usepackage{rotating}
\usepackage{subfigure}
\usepackage{color}
\usepackage{blindtext}
\usepackage{lipsum}
\usepackage{multirow}
\usepackage{physics}
\usepackage{orcidlink}

	\allowdisplaybreaks
\begin{document}
	
	
	\title{\textcolor{BlueViolet}{Hybrid stars with hyperons: structure based on QCD sum rule coupling constants}}

	\author{F.~Moradi Jangal$^{a}$\orcidlink{0009-0009-9049-0528}}
\email{F.Moradi71@ut.ac.ir }

\author{H.~R.~Moshfegh$^{a,b}$\orcidlink{0000-0002-9657-7116}}
\email{hmoshfegh@ut.ac.ir }
\thanks{Corresponding author}

\author{K.~Azizi$^{a,c}$\orcidlink{0000-0003-3741-2167}}
\email{kazem.azizi@ut.ac.ir}

\affiliation{
	$^{a}$Department of Physics, University of Tehran, North Karegar Avenue, Tehran 14395-547, Iran\\		
	$^{b}$	Centro Brasileiro de Pesquisas Fısicas, Rua Dr. Xavier Sigaud,
	150, URCA, Rio de Janeiro CEP 22290-180, RJ, Brazil	\\	
	$^{c}$Department of Physics, Dogus University, Dudullu-\"Umraniye, 34775 Istanbul, T\"urkiye
	}

\date{\today}

\preprint{}
	\begin{abstract}
	
		We present a comprehensive study of hybrid stars composed of hadrons, leptons, and quarks within a relativistic mean-field framework. Using coupling constants derived from QCD sum rules (QCDSR), we first determine the bulk properties of nuclear matter and evaluate the single-particle potentials of nucleons and hyperons to constrain the hadronic sector.
		The equation of state (EOS) under beta equilibrium is then constructed employing the  $\sigma$-$\omega$-$\rho$ model for the hadronic phase, while the quark phase is described using both the MIT bag model and the Nambu–Jona-Lasinio (NJL) model. The hadron–quark phase transition is analyzed through both Gibbs and Maxwell constructions. Based on resulting EOSs, we obtain the mass–radius relations of hybrid stars, investigate particle fractions and their radial distributions, and calculate the tidal Love number ($\mathcal{K}_{2}$) and the dimensionless tidal deformability ($\varLambda$). Our results provide quantitative predictions relevant for comparison with current multimessenger astrophysical observations.

	\end{abstract}
	
	
	\maketitle
	
	\renewcommand{\thefootnote}{\#\arabic{footnote}}
	\setcounter{footnote}{0}
	
	\section {Introduction}\label{sec:one}

	In recent decades, the study of matter under extreme conditions of density and temperature has become a central topic in modern nuclear physics and astrophysics. Advances in nuclear many-body theory have provided important insights into the behavior of strongly interacting matter, particularly in regimes inaccessible to terrestrial experiments. Compact stars, and especially neutron stars (NSs), serve as natural laboratories for probing such extreme environments, where gravitational, strong, weak, and electromagnetic interactions collectively determine the macroscopic structure of the star.
	The existence of neutron stars was first proposed by Baade and Zwicky in 1934 \cite{Baade:1934wuu}. Since then, numerous theoretical approaches have been developed to describe their structure and physical properties \cite{Chin:1974sa, Huber:1997mg, Heiselberg:1999mq, Lattimer:2000nx, Weber:2004kj}. A fundamental ingredient in these studies is the equation of state (EOS) of dense matter, which determines the mass–radius relation and constrains the internal composition of neutron stars. Despite significant theoretical progress, uncertainties in the behavior of matter at supra‑nuclear densities continue to limit our understanding of their maximum mass and microscopic structure. Therefore, precise astrophysical observations play a crucial role in constraining realistic EOSs \cite{Ozel:2016oaf}. Observational breakthroughs have considerably tightened these constraints. The Shapiro delay measurement of PSR J1614–2230 \cite{Demorest:2010bx}, subsequent timing analyses by NANOGrav \cite{NANOGrav:2017wvv, NANOGrav:2019jur}, the mass and radius determination of PSR J0740+6620 using NICER and XMM-Newton \cite{Fonseca:2021wxt}, the discovery of the massive pulsar PSR J0952–0607 \cite{Romani:2022jhd}, and NICER timing observations of PSR J0537–6910 \cite{Ho:2015vza,Ho:2020vxt} have established the existence of neutron stars with masses around or above $2M_{\odot}$. Furthermore, gravitational-wave detections from binary neutron star mergers, such as GW170817 \cite{LIGOScientific:2018hze} and GW190425 \cite{LIGOScientific:2020aai}, have provided independent constraints through the tidal deformability parameter \cite{Flanagan:2007ix,Hinderer:2007mb}. In particular, GW170817 constrains the dimensionless tidal deformability to $\varLambda_{1.4} = 190_{-120}^{+390}$ and restricts the radius of a $1.4M_{\odot}$ neutron star to 11.82–13.72 km \cite{LIGOScientific:2018cki, Lim:2019som,Malik:2018zcf}. These multimessenger observations define stringent empirical boundaries for viable dense matter models \cite{Raaijmakers:2019qny, Miller:2021qha, Chatziioannou:2020pqz, Li:2021crp}.
	At densities exceeding approximately $(2$–$3)\rho_{0}$, additional degrees of freedom beyond nucleons are expected to appear \cite{Schulze:1998jf,Djapo:2008au,Vidana:2015rsa,Lonardoni:2013gta}. In particular, the emergence of hyperons in the inner core is energetically favored. However, the inclusion of hyperons generally softens the EOS and often leads to maximum masses below the $\sim 2~M_{\odot}$ observational limit \cite{Demorest:2010bx, NANOGrav:2017wvv, NANOGrav:2019jur}. This tension, widely known as the hyperon puzzle \cite{Looee:2025dgx}, remains one of the major open issues in nuclear astrophysics.	
	In the present work, we construct the hadronic equation of state within the relativistic $\sigma$–$\omega$–$\rho$ framework \cite{Chin:1974sa, Mueller:1996pm, Walecka:1974qa, Uechi:2006pz}. A distinctive feature of our approach is that the baryon–meson coupling constants are not treated as purely phenomenological free parameters. Instead, they are determined from QCD sum rules, thereby establishing a direct connection between the underlying quark–gluon dynamics and the macroscopic properties of dense matter. This strategy reduces the arbitrariness in the choice of couplings and provides a microscopic foundation for the hadronic sector.
	QCD sum rule inspired couplings have previously been employed to investigate hyperonic nuclear matter and neutron star properties. In particular, in Ref.~\cite{Jangal:2025maa} these couplings were applied to hyperonic neutron stars while supporting massive stellar configurations. Independent studies have also explored similar QCD-based coupling schemes in dense matter applications (see, e.g., Ref.~\cite{Colucci:2013pya}), further confirming their phenomenological viability and consistency with astrophysical observations. Previous investigations have extensively explored hybrid stars based on relativistic
	mean-field hadronic models combined with both the MIT bag model and the
	three-flavor NJL model using Gibbs and Maxwell phase-transition constructions
	\cite{Bonanno:2011ch,Shahrbaf:2019vtf,Shahrbaf:2020uau,Contrera:2016phj}.
	In particular, Ref.~\cite{Karimi:2022vxw} investigated hybrid stars employing the
	 $\sigma$–$\omega$–$\rho$ model with hyperons together with the MIT bag
	and NJL quark models, considering both Gibbs and Maxwell constructions and
	their corresponding mass--radius relations. In this work, we employed QCD sum rule coupling constants for baryons, which are reliable, have been tested by other groups, and provide precise results for observable quantities.
	Nevertheless, a systematic hybrid star analysis incorporating both Maxwell and Gibbs phase transition constructions within this QCD-motivated coupling framework has not yet been fully addressed. 
	At higher densities, we also consider the possibility of a transition to deconfined quark matter, leading to hybrid star configurations. To explore this scenario, two different quark matter descriptions are employed, and the hadron–quark phase transition is modeled using both Maxwell and Gibbs constructions. This comprehensive treatment allows us to examine the sensitivity of neutron star properties to the phase transition mechanism and to the quark matter model within a consistent microscopic scheme.
	Our results demonstrate that, within this extended hybrid framework and under QCD sum rule constrained couplings, stable configurations with maximum masses exceeding $2~M_{\odot}$ can be obtained even in the presence of hyperons. This finding indicates that the hyperon puzzle can be alleviated within a framework where the microscopic input is anchored in QCD-based calculations, while remaining consistent with current observational constraints.
	The paper is organized as follows. In Sec.\ref{sec:two}, we present the theoretical formalism and the construction of the hadronic and quark equations of state, as well as the phase transition prescriptions. Sec.\ref{sec:three} is devoted to the numerical results and their comparison with observational data. Finally, our conclusions are summarized in Sec.~\ref{sec:four}.

\section {Theoretical Framework}
\label{sec:two}
	\subsection{Lagrangian Density}
We will employ a Lagrangian to characterize the dynamics of particles within the core of Neutron star. The Lagrangian density includes components for leptons, baryons, and mesons, and their interaction terms. It is represented as follows \cite{Glendenning:1984jr}:
	\begin{equation}
		\mathscr{L} = \sum_{B} \mathscr{L}_{B} + \sum_{\mathcal{M}} \mathscr{L}_{\mathcal{M}} +\sum_{\lambda} \mathscr{L}_{\lambda} + \mathscr{L}_{int}.
	\end{equation}
	where B, $\mathcal{M}$, and $\lambda$ denote baryons, mesons, and leptons, respectively.
	The Lagrangian for hadronic sector is described within the relativistic mean‑field (RMF) framework based on the exchange of three mesons: the scalar–isoscalar $\sigma$, the vector–isoscalar $\omega$, and the vector–isovector $\rho$. Within this $ \sigma,\omega, \rho $ model, the Lagrangian density takes the form:
	
	\begin{equation}
		\label{eqn:L2}
		\begin{split}
			\mathscr{L}  = & \sum_{B} \bar{\psi}_{B} (i \gamma_{\mu} \partial^{\mu} - m_{B}
			+ g_{BB\sigma}\sigma - g_{BB\omega}\gamma_{\mu}\omega^{\mu} - \dfrac{1}{2}  g_{BB\rho}\gamma_{\mu} 	\boldsymbol{\tau}.\boldsymbol{\rho}^{\mu}) \psi_{B} \\
			& + \dfrac{1}{2}(\partial_{\mu}\sigma\partial^{\mu}\sigma - m^{2}_{\sigma}\sigma^{2}) - \dfrac{1}{4} \omega_{\mu\nu} \omega^{\mu\nu} + \dfrac{1}{2} m^{2}_{\omega} \omega_{\mu} \omega^{\mu}\\
			& -\dfrac{1}{4} \boldsymbol{\rho}_{\mu\nu} . \boldsymbol{\rho}^{\mu\nu} + \dfrac{1}{2} m^{2}_{\rho} \boldsymbol{\rho}_{\mu}.\boldsymbol{\rho}^{\mu} - \dfrac{1}{3} bm_{n} (g_{NN\sigma}\sigma)^{3} - \dfrac{1}{4} c (g_{NN\sigma}\sigma)^{4} \\
			& + \sum_{\lambda} \bar{\psi}_{\lambda}(i \gamma_{\mu} \partial^{\mu} - m_{\lambda})\psi_{\lambda}.\\
		\end{split}
	\end{equation}
	In this context, we consider baryons to be composed of neutrons, protons, and hyperons, while mesons include the $\sigma$, $\omega$, and $\rho$ particles. Leptons are represented by electrons and muons. In the Lagrangian density, $\partial^{\mu}$ denotes the four-derivative operator, and $\gamma_{\mu}$ are the covariant Dirac matrices. The tensors $\boldsymbol{\rho}_{\mu\nu}$ and $\omega_{\mu\nu}$ correspond to the field strength tensors of the $\rho$ and $\omega$ mesons, respectively. The spinor fields are indicated by $\psi$, with their adjoint spinors defined as $\bar{\psi} \equiv \psi^{\dagger} \gamma_0$. The coupling constants in the Lagrangian density, labeled as $g_{BB\sigma}$, $g_{BB\omega}$, $g_{BB\rho}$, along with parameters b and c, represent the interactions between baryons and mesons, as well as scalar self-interactions (b and c). Furthermore, $\boldsymbol{\tau}$ denotes the isospin matrices.
    In this study, a relativistic mean-field approximation (RMFA) is used to examine baryonic matter at high densities. To describe how the energy functional depends on density, nonlinear interactions among the fields are introduced, drawing on concepts from effective field theory \cite{Mueller:1996pm}. Specifically, the focus is on various types of nonlinearities involving the scalar-isoscalar ($\sigma$), vector-isoscalar ($\omega$), and vector-isovector ($\rho$) fields. Within the RMFA approach, it is assumed that particles do not directly interact with one another but instead experience an average influence from their surrounding environment.
	\subsection{Chemical Potential and Meson Fields}
As highlighted by Glendenning \cite{Glendenning:1984jr}, the issue of chemical equilibrium can be tackled without focusing on specific reactions. In the case of frozen static matter, the chemical potential ($\mu$) corresponds to the energy of a particle at the Fermi surface. By solving the Euler-Lagrange equations and applying the mean-field approximation to the meson fields, we can find the fields and their eigenvalues. This enables us to represent the baryon fields in momentum space as:
	\begin{equation}
		[\gamma_{\mu}(k^{\mu}-g_{BB\omega}\omega^{\mu}-\dfrac{1}{2}g_{BB\rho}\boldsymbol{\tau}.\boldsymbol{\rho}^{\mu})-(m_{B}-g_{BB\sigma}\sigma)]\psi_{B}(k)=0,
	\end{equation}
    and the eigenvalues for particles and antiparticles are:
	\begin{equation}
		\label{Eq:ch2}
		e_{B}(k) = g_{BB\omega}\omega_{0} + g_{BB\rho}\rho_{03}I_{B3} + \sqrt{k^{2}+(m_{B}-g_{BB\sigma}\sigma)^{2}},
	\end{equation}
	\begin{equation}
		\bar{e}_{B}(k) = -g_{BB\omega}\omega_{0} - g_{BB\rho}\rho_{03}\bar{I}_{B3} + \sqrt{k^{2}+(m_{B}-g_{BB\sigma}\sigma)^{2}}.
	\end{equation}
Here, \( I_{B3} \) denotes the isospin projection of baryon \( B \). Furthermore, the meson fields \( \sigma \), \( \omega_0 \), and \( \rho_{03} \) in uniform static matter can be expressed as:
	\begin{equation}
		\label{eqn:m-f-one}
		\omega_{0} = \sum_{B} \dfrac{g_{BB\omega}}{m_{\omega}^{2}}\rho_{B},
	\end{equation}
	\begin{equation}
		\label{eqn:m-f-two}
		\rho_{03} = \sum_{B} \dfrac{g_{BB\rho}}{m_{\rho}^{2}}I_{B3}\rho_{B},
	\end{equation}
	\begin{equation}
		\label{eqn:m-f-tree}
		m_{\sigma}^{2}\sigma = - b m_{N}g_{NN\sigma}(g_{NN\sigma}\sigma)^{2} - c g_{NN\sigma}(g_{NN\sigma}\sigma)^{3}\\
		+\sum_{B}\dfrac{2J_{B}+1}{2\pi^{2}} g_{BB\sigma}\int_0^{k_{B}} \dfrac{m_{B}-g_{BB\sigma}\sigma}{\sqrt{k^{2}+(m_{B}-g_{BB\sigma}\sigma)^{2}}}k^{2}dk.
	\end{equation}
	
	The spin of a baryon is represented by \( J_{B} \). The overall baryon number density, denoted as \( \rho_{B} \), is connected to the Fermi momentum \( k_{F} \) through the following equation:
	\begin{equation}
		\rho_{total} = \sum_{i} \rho_{i} = \sum_{i} \dfrac{k_{F_{i}}^{3}}{3\pi^{2}},
	\end{equation}
	where, $i$ is the set of baryons in the model.
	\subsection{Equation of State}
	The equation of state (EOS) of dense matter is obtained from the energy–momentum tensor derived from the Lagrangian density given in Eq. \ref{eqn:L2}. Within the relativistic mean‑field approximation, the total energy density consists of contributions from baryons, leptons, and meson fields.
The total energy and pressure  density can be written as:
	\begin{equation}
		\begin{split}
			\varepsilon =
			&\sum_{B}\dfrac{2J_{B}+1}{2\pi^{2}}\int_0^{k_{B}} \sqrt{k^{2}+(m_{B}-g_{BB\sigma}\sigma)^{2}}k^{2}dk \\
			&+\sum_{\lambda}\dfrac{1}{\pi^{2}}\int_0^{k_{\lambda}}\sqrt{k^{2}+m_{\lambda}^{2}}k^{2}dk\\
			& +\dfrac{1}{3} b ~ m_{N}(g_{NN\sigma}\sigma)^{3} +\dfrac{1}{4} c (g_{NN\sigma}\sigma)^{4}+ \dfrac{1}{2}m_{\sigma}^{2}\sigma^{2} + \dfrac{1}{2}m_{\omega}^{2}\omega_{0}^{2} + \dfrac{1}{2}m_{\rho}^{2}\rho_{03}^{2}, \\
		\end{split}
	\end{equation}
	\begin{equation}
		\begin{split}
			p = &\dfrac{1}{3}\sum_{B}\dfrac{2J_{B}+1}{2\pi^{2}}\int_0^{k_{B}}\dfrac{k^{4}dk}{ \sqrt{k^{2}+(m_{B}-g_{BB\sigma}\sigma)^{2}}}\\
			&+\dfrac{1}{3}\sum_{\lambda}\dfrac{1}{\pi^{2}}\int_0^{k_{\lambda}}\dfrac{k^{4}dk}{\sqrt{k^{2}+m_{\lambda}^{2}}}\\
			& -\dfrac{1}{3} b ~ m_{N}(g_{NN\sigma}\sigma)^{3} -\dfrac{1}{4} c (g_{NN\sigma}\sigma)^{4}- \dfrac{1}{2}m_{\sigma}^{2}\sigma^{2} + \dfrac{1}{2}m_{\omega}^{2}\omega_{0}^{2} + \dfrac{1}{2}m_{\rho}^{2}\rho_{03}^{2}. \\
		\end{split}
	\end{equation}
	
	Here, $ m_{N} $ and $ m_{B}$ represent the masses of the nucleon and baryons, respectively. The meson masses are indicated by $m_{\sigma}$, $m_{\omega}$, and $m_{\rho}$. This research uses a model of a cold neutron star composed of neutrons, protons, electrons, muons, and hyperons  ($\Xi^{-}$ and $\Lambda$). Because of the large rest mass of the tau lepton, its contribution is regarded as negligible. The possible processes to establish the beta equilibrium condition in the system are as follows:
	\begin{equation}
		\label{U1}
		n \longrightarrow p + e + \bar{\nu}_{e},
	\end{equation}
	and
	\begin{equation}
		\label{U2}
		p + e \longrightarrow n + \nu_{e}.
	\end{equation}
 When the energy of electrons increases enough, muons can be produced as,
	\begin{equation}
		e \longrightarrow \mu + \bar{\nu}_{\mu} + \nu_{e}.
	\end{equation}

	At higher densities, $\Lambda$ and  $\Xi^{-}$ could be produced via the following processes,
	
	\begin{equation}
		\label{eq:Lamb}
		n + n \longrightarrow n + \Lambda.
	\end{equation}
		and
	\begin{equation}
		\label{eq:Sig}
		n + e \longrightarrow \Xi^{-} + \nu_{e},
	\end{equation}
	The thermodynamic equilibrium conditions for chemical potentials that govern the above equations are as follows:
	\begin{equation}
		\label{eqn:mu}
		\mu_{i} = B_{i} \mu_{n} - Q_{i} \mu_{e}.
	\end{equation}	
	Where, $B_{i}$, and $Q_{i}$ denote the baryon charge and electric charge of each species. For Eqs. (\ref{U1}) and (\ref{U2}), the chemical potential of neutrinos is zero since they escape the star as it evolves. 
	Additionally, the system maintains charge neutrality, which is established internally, as:
	\begin{equation}
		\label{eqn:ch-n}
		\rho_{p} = \rho_{e} + \rho_{\mu} + \rho_{\Xi^{-}},
	\end{equation}
	
	where $\rho_{i}$ is the number density of each particle. Finally, number density conservation is expressed as:
	\begin{equation}
		\label{eqn:b-d}
		\rho_{B} = \rho_{n} + \rho_{p} + \rho_{\Xi^{-}} +  \rho_{\Lambda}.
	\end{equation}

	\subsection{Phase transition and Mixed phase}
		
In this work, we investigate the hadron--quark phase transition using both the Maxwell and Gibbs constructions, which represent two limiting scenarios for first-order phase transitions in dense matter.

Within the Maxwell construction, each phase is required to satisfy local electric charge neutrality independently. The phase transition occurs when the baryon chemical potential and the pressure of the hadronic ($H$) and quark ($Q$) phases become equal, namely  \cite{Sen:2022qol}
\begin{equation}
	P_H = P_Q ,
	\qquad
	\mu_B^H = \mu_B^Q .
	\label{eq:Maxwell_condition}
\end{equation}
In this case, the electron chemical potential is generally discontinuous across the transition,
\begin{equation}
	\mu_e^H \neq \mu_e^Q ,
\end{equation}
which implies the presence of a sharp interface between the two phases. Consequently, no spatially extended mixed phase (MP) appears, and the transition is characterized by a constant-pressure plateau in the equation of state.

In contrast, the Gibbs construction allows for global electric charge neutrality of the system, while the individual phases may carry nonvanishing electric charge. Under these conditions, both independent chemical potentials remain continuous across the transition. The equilibrium conditions are given by
\begin{equation}
	P_H = P_Q =  P_{MP} ,
	\qquad
	\mu_B^H = \mu_B^Q ,
	\qquad
	\mu_e^H = \mu_e^Q .
	\label{eq:Gibbs_condition}
\end{equation}
As a result, a mixed phase region emerges over a finite density interval, where hadronic and quark matter coexist. In this region, the total baryon density can be expressed as
\begin{equation}
	\rho_B = (1-\chi)\rho_B^H + \chi \rho_B^Q ,
\end{equation}
where $\chi$ denotes the volume fraction of quark matter and varies continuously from 0 to 1. Essentially, the $\chi$ parameter delineates three distinct phases, as described below:

\textbf{$\bullet$} Hadronic phase  \hspace{5em} $\chi$  $=$  0

\textbf{$\bullet$} Mixed phase   \hspace{5em}   0 $<$ $\chi$  $<$  1

\textbf{$\bullet$} Pure quark phase  \hspace{4em} $\chi$  $=$  1

In the mixed phase, the principle of global conservation is employed \cite{Glendenning:1992vb}. Consequently, both the baryon number density and charge neutrality can be expressed as functions of $\chi$,

\begin{equation}
	(1 - \chi)\rho_H + \chi\rho_Q = \rho_B,
\end{equation}
\begin{equation}
	(1 - \chi)q_H + \chi q_Q = 0.
\end{equation}

In the present analysis, both constructions are implemented in order to examine the robustness of the stellar configurations against the assumed phase transition mechanism and to assess how the choice of construction influences the maximum mass, radius, and internal composition of hybrid stars.

	\subsection{Quark phase}

	\subsubsection{MIT bag model}
We begin by providing an overview of the MIT bag model \cite{Ghose:1975ib}. The thermodynamic potential for quarks of flavor f = u, d, s  at zero temperature is expressed as follows:

	\begin{equation}
		\begin{aligned}
				\label{eqn:Omegaa}
			\Omega_f(\mu_f) &= -\frac{1}{4\pi^2} \Bigg[
			\mu_f \left( \mu_f^2 - \frac{5}{2} m_f^2 \right) \sqrt{\mu_f^2 - m_f^2}
			+ \frac{3}{2} m_f^4 \ln \left( \frac{\mu_f + \sqrt{\mu_f^2 - m_f^2}}{m_f} \right)
			\Bigg] \\
			&\quad + \frac{\alpha_c}{2\pi^3} \Bigg[
			3 \left(
			\mu_f \sqrt{\mu_f^2 - m_f^2}
			- m_f^2 \ln \left( \frac{\mu_f + \sqrt{\mu_f^2 - m_f^2}}{m_f} \right)
			\right)^2 \\
			&\qquad - 2 (\mu_f^2 - m_f^2)^2
			+ 3 m_f^4 \ln \left( \frac{m_f}{\mu_f} \right)^{2} \\
			&\qquad + 6 \ln \left( \frac{\xi}{\mu_f} \right)
			\Big[
			\mu_f m_f^2  (\mu_f^2  - m_f^2)^{1/2}
			- m_f^4 \ln \left( \frac{\mu_f + \sqrt{\mu_f^2 - m_f^2}}{m_f} \right)
			\Big]
			\Bigg].
		\end{aligned}
	\end{equation}

	The first line of Equation (\ref{eqn:Omegaa}) corresponds to the kinetic term, whereas the subsequent lines represent the one-gluon-exchange interaction term, which is proportional to the QCD fine-structure constant $\alpha_c$ \cite{Ghose:1975ib}. Here, $\mu_f$ and $m_f$ denote the chemical potential and mass of the quark with flavor $f$, respectively, and the renormalization scale $\xi$ is set to 313 MeV  accordance with \cite{Khanmohamadi:2019jky}. Furthermore, the thermodynamic potential $\Omega$ is defined as:
	
	\begin{equation}
\Omega = \sum_f \Omega_f + B.
	\end{equation}
	
The parameter B, known as the bag constant, is treated as a free variable within the model. Fundamentally, B is defined as the difference in energy density between the perturbative vacuum and the true vacuum \cite{Aziz:2019rgf}. Consequently, the number density, pressure, and total energy density can be derived as follows:

\begin{equation}
n_f = - \frac{\partial \Omega}{\partial \mu_f},
\end{equation}

\begin{equation}
P = - \Omega,
\end{equation}

\begin{equation}
\epsilon = \Omega + \sum_f \mu_f n_f.
\end{equation}

\subsubsection{The NJL model}

This section presents the three-flavor formulation of the Nambu–Jona-Lasinio (NJL) model. The Lagrangian most frequently employed within this framework is expressed as \cite{Rehberg:1995kh},

\begin{equation}
	L=\bar{q}(i\slashed{\partial}-\hat{m})q+L_{sym}+L_{det}.
\end{equation}

$ q=(u,d,s)^{T} $ denote a quark field comprising three distinct flavors, and let the associated quark mass matrix be given by  $ \hat{m}=diag (m_{u},m_{d},m_{s}) $. The Lagrangian incorporates two independent interaction terms, which are specified by:

\begin{equation}
	L_{sym}=G\sum_{a=0}^{8} \left[(\bar{q}\lambda_{a}q)^{2} + (\bar{q}  i\gamma_{5}\lambda_{a}q )^{2}\right],
\end{equation}
and 
\begin{equation}
	L_{det} = -K \left[det \Bigg(\bar{q}(1+\gamma_{5})q \Bigg)+ det\Bigg(\bar{q}(1-\gamma_{5})q\Bigg) \right].
\end{equation}

In this context,  G  and  K  denote the coupling constants of the theoretical framework. The term $ L_{sym} $ represents a four-point interaction that is symmetric under the $ U(3)_{L} \times U(3)_{R} $ group, where the matrices $ \lambda_{a} $ for $ a=1,\dots,8 $ serve as the generators of the $ SU(3) $ group.  In addition, we define
$\lambda_0=\sqrt{\frac{2}{3}}\,I,$
where $I$ is the $3\times3$ identity matrix in flavor space. Within flavor space, the term  $ L_{det} $, associated with the 't Hooft interaction, is expressed as a determinant and constitutes a six-point interaction characterized by maximal flavor mixing. Notably, $ L_{det} $ is $SU(3)_{L} \times SU(3)_{R}$ symmetry and breaks the $ U(1) $ symmetry, while $ U(1) $ is unbroken by $ L_{sym} $ \cite{Buballa:2003qv}. Furthermore, the quark self-energy in the Nambu–Jona-Lasinio (NJL) model leads to the following gap equation:

\begin{equation}\label{m1}
	M_{i} =m_{i}-4G\varphi_{i} +2K\varphi_{j}\varphi_{k}.
\end{equation}
Where $ M $ denote the constituent quark mass, $ (i,j,k) $ represent any permutation of the quark flavors  $ (u,d,s) $, and $\varphi_{i} = <\bar{q}_{i}q_{i}>$ corresponds to the quark condensate parameter associated with flavor $i$ \cite{Khanmohamadi:2019jky}.

Divergent integrals arise within the Nambu–Jona-Lasinio (NJL) model, necessitating the implementation of a regularization procedure. Various regularization techniques are available; however, for thermodynamic applications, the use of a three-momentum cutoff—either sharp or smooth—is predominantly advocated. In the present work, we adopt a sharp three-momentum cutoff scheme. The cutoff parameter, denoted by $ \Lambda_{c} $, constitutes one of the five fundamental parameters of the NJL model. The remaining four parameters include the coupling constants  $K$  and  $G$, as well as the bare quark masses $ m_{u} $,  $m_{d} $ and $ m_{s}$. These parameters are calibrated by fitting to five empirical observables: the masses of the pseudoscalar mesons  $\eta^{'}$, $\eta$ and $K$, the pion mass  $ m_{\pi}$, and the pion decay constant $f_{\pi}$. We use three distinct parameter sets employed in the literature. The RKH parameter set corresponds to the fits by Rehberg, Klevansky, and Hufner \cite{Rehberg:1995kh}, the HK set pertains to the fits by Hatsuda and Kunihiro \cite{Hatsuda:1994pi}, and the LKW set is derived from the fits by Lutz, Klimt, and Weise \cite{lutz:1}.

Within this framework, at zero temperature, the mean-field thermodynamic potential is characterized by the following expression:

\begin{align}
	\Omega (\mu_{f},\varphi_{f}) = &\sum_{f=u,d,s} \Omega_{M_{f}}(\mu_{f}) + 2G (\varphi_{u}^{2} + \varphi_{d}^{2}+ \varphi_{s}^{2})\notag\\&-4K \varphi_{u} \varphi_{d} \varphi_{s} + \Omega_{0}.
\end{align}
Where $ \Omega_{M_{f}} $ is defined in Eq. (\ref{equa1}) and represents the contribution from a gas of quasiparticles possessing mass $ M_{f} $.
 $ \Omega_{M_{f}}$ is given by the below expression ,at zero temperature,
 
\begin{equation}\label{equa1}
	\Omega_{M_{f}}(\mu_{f}) = \frac{-N_{c}}{\pi^{2}}  \int_{P_{F,f}}^{\Lambda}  E_{p,f} p^{2} \mathrm{d}p - \mu_{f} n_{f}.
\end{equation}

Where $ n_{f} = \frac{(P_{F,f})^3}{\pi^{2}} $, $ E_{p,f} = \sqrt{p^{2} + M_{f}^{2}} $, and $ P_{F,f} = \sqrt{\mu_{f}^{2} - M_{f}^{2}} $ corresponds to the number density, the on-shell energy, and the Fermi momentum associated with quarks of flavor $ f $. Also $\Lambda$ is a sharp three-momentum cutoff, and $ N_{c}=3 $ is the number of colors, respectively. The quark condensates  are determined by minimizing $\Omega$ at its stationary points $ (\delta \Omega/\delta\varphi_{f}=0) $. Consequently, the quark condensates can be written as follows: 
\begin{equation}
	\varphi _{f} = \frac{-N_{c}}{\pi^{2}} \int_{P_{F,f}}^{\Lambda} \frac{M_{f}}{E_{p,f}} p^{2}\mathrm{d}p.
\end{equation}
The equations $ (\varphi_{f}) $ must be solved self-consistent manner alongside Eq. (\ref{m1}), creating a group of three coupled gap equations for the constituent masses. Additionally, the parameter $ \Omega_{0} $ can be determined easily by imposing the condition that the pressure vanishes in the limit as $ \mu, T \rightarrow 0 $. After resolving these self-consistent equations, other thermodynamic quantities can subsequently be derived:
\begin{equation}
	P = -\Omega, \qquad \qquad \epsilon = \Omega + \sum_{f} \mu_{f} n_{f}, \qquad\qquad n_{f}=-\frac{\partial \Omega}{\partial \mu_{f}}.
\end{equation}

\subsubsection{Electrons and Beta Equilibrium} 

The weak decay process $ (d \leftrightarrow u + e + \bar{\nu}_{e} \leftrightarrow s ) $ occurring within the quark matter results in the generation of electrons in this medium. Given the presence of quarks, the electron mass can be regarded as negligible; consequently, electrons are modeled as a massless, non-interacting fermionic gas. And consequently,

\begin{equation}
	P_{e}= \frac{\mu_{e}^{4}}{12 \pi^{2}}, \qquad\qquad \epsilon_{e}= \frac{\mu_{e}^{4}}{4 \pi^{2}}, \qquad\qquad n_{e}=\frac{\mu_{e}^{3}}{3 \pi^{2}}.
\end{equation}
and,

\begin{equation}
	P_{tot} = P + P_{e},\qquad\qquad \epsilon_{tot}=\epsilon+\epsilon_{e}.	
\end{equation}

Furthermore, within beta-stable matter, the chemical potentials of quarks and electrons can be represented as follows:

\begin{equation}
	\mu_{d}=\mu_{s}, \qquad\qquad \mu_{d}= \mu_{u}+\mu_{e}.
\end{equation}

The condition of charge neutrality in quark matter leads to:
\begin{equation}
	0=\frac23 n_{u}-\frac13 (n_{d}+n_{s})-n_{e},
\end{equation}
and the baryon number density is:
\begin{equation}
	\rho_{B}=\frac13 (n_{u}+n_{d}+n_{s}).
\end{equation} 

The parameters used in the MIT Bag and NJL models are summarized in Table~\ref{tab:model_parameterss}.

\begin{table}[H]
	\centering
	\caption{Parameters used in the MIT Bag and NJL models.}
	\label{tab:model_parameterss}
	\begin{center}
	
	\begin{tabular}{|| c || c | c| c| c ||}
		\hline
		Parameter & MIT & RKH & HK & LKW \\
		\hline\hline
		$m_u$ (MeV) & 2.2 & 5.32 & 5.32 & 3.55 \\
		$m_d$ (MeV) & 4.75 &5.32 & 5.32 & 3.55\\
		$m_s$ (MeV) & 95.5 & 140 & 135.5 & 87\\
		$\alpha_c$ & 0.3 & -- & -- & -- \\
		$\xi$ (MeV) & 313 & -- & -- & -- \\
		$\Lambda_c$ (MeV) & -- & 602.3& 631.4 & 750 \\
		$G\Lambda_c^2$ & -- & 1.835 & 1.835 & 1.820 \\
		$K\Lambda_c^5$ & -- & 12.36 & 9.29 & 8.90 \\
		\hline
	\end{tabular}
		\end{center}
\end{table}

\section {Results and discussion}\label{sec:three}	
\label{sec:result}

A central feature of the present work is the use of QCD sum rule (QCDSR)–derived meson–baryon coupling constants in the hadronic sector of the equation of state. Conventional RMF studies typically determine these couplings by fitting to selected nuclear observables—such as saturation properties, symmetry energy, and hypernuclear potential depths. While this phenomenological approach is effective, it does not provide a direct connection to the underlying theory of strong interactions.
The QCD sum rule approach is a nonperturbative method that relates hadronic observables to QCD vacuum condensates through the operator product expansion (OPE) and dispersion relations. 
In this framework, correlation functions of interpolating currents are computed in terms of quark and gluon degrees of freedom, and matched to a phenomenological representation involving hadron properties \cite{Azizi:2016dhy, Doi:2003cd}.

By considering appropriate baryonic interpolating fields and selecting Lorentz structures sensitive to meson–baryon interactions, one can extract relations between the coupling constants and QCD condensates \cite{Azizi:2016dhy, Doi:2003cd,Azizi:2015bxa}. 
The procedure typically involves constructing two-point correlation functions with an external mesonic current, performing the OPE up to a given dimension, and applying Borel transformations to suppress higher-order contributions and continuum effects. 
The resulting sum-rule expressions connect the meson–baryon coupling constants with quark condensates, gluon condensates, and mixed condensates evaluated in the QCD vacuum. 
In the present study, the hadronic sector is restricted to $\Lambda$ and $\Xi^-$. The exclusion of  $\Sigma$ hyperons is motivated by two key considerations: experimental evidence regarding hyperon–nucleus potentials and methodological consistency with the QCD sum rule framework.

First, recent experimental data from hypernuclear spectroscopy and ($pion^{-}$ $K^{+}$) reactions provide strong evidence that the $\Sigma$-nucleus interaction is significantly repulsive. In contrast to the well-established attractive potentials for $\Lambda$ and $\Xi^{-}$  hyperons, the $\Sigma$ potential at saturation density is estimated to be +20 to +30 Mev \cite{Schaffner_Bielich_2000}. Within the context of beta-equilibrated neutron star matter, such a repulsive potential substantially raises the energy threshold for the appearance of $\Sigma^-$ and $\Sigma^0$  hyperons. Consequently, their onset is shifted to densities significantly higher than those typically reached in the cores of stable neutron stars, rendering their impact on the equation of state (EOS) negligible compared to $\Lambda$ and $\Xi^-$. Second, we prioritize methodological consistency within our QCDSR approach. While various sets of coupling constants (CCs) for the hyperon sector exist in the literature, those pertaining to $\Sigma$  hyperons are often not derived directly from pure QCDSR calculations. Instead, the only reported $\Sigma$ coupling values \cite{Erkol:2006eq} are inferred using symmetry considerations (such as SU(3) flavor symmetry) to supplement missing or unstable sum-rule results. In this work, we exclusively employ coupling constants that have been directly calculated via the QCDSR method, without recourse to secondary symmetry assumptions. This ensures that our hadronic interaction remains strictly rooted in the nonperturbative QCD condensate expansion. Since the $\Lambda$ and $\Xi^-$  couplings are both directly accessible through QCDSR and represent the most probable strange degrees of freedom to appear in dense matter, we focus on these species to maintain a robust and microscopically consistent description of the hybrid star interior.

The meson--baryon coupling constants employed in this work are summarized in Table~\ref{table:c.c}. 
\begin{table}[H]
	\caption{QCDSR Coupling constant.}
	\label{table:c.c}
	\small
	\begin{center}
		\begin{tabular}{|| c | c | c | c | c | c | c | c | c | c ||}
			\hline
			$g_{NN\sigma}$ & $g_{NN\omega}$ & 	$g_{NN\rho}$& 	$g_{\Lambda\Lambda\sigma}$ & 	$g_{\Lambda\Lambda\omega}$  & 	$g_{\Lambda\Lambda\rho}$  &	$g_{\Xi^{-}\Xi^{-}\sigma}$ & 	$g_{\Xi^{-}\Xi^{-}\omega}$ & 	$g_{\Xi^{-}\Xi^{-}\rho}$  & Reference \\ 
			\hline\hline
			8$\pm$ 2.0 & 8.9$\pm$ 1.5& 5.9$\pm$ 1.3& 7$\pm$ 1.9& 7.1$\pm$ 1.1& 0.0& 1.9$\pm$ 0.4& 1.5$\pm$ 1.1& 4.2$\pm$ 2.1& Refs. \cite{Aliev_2007,Aliev:2009ei, Wang:2007yt, Erkol:2006sa, Zamiralov:2013gva}  \\
				\hline
					\multicolumn{10}{||c||}{Selected coupling constants used in this work} \\
				\cline{1-10}
				\hline\hline
				8.61&8.90 &7.20 &5.70 &6.45 & 0.0 &1.50 &1.55 &6.00 & This work \\
			
			\hline
		\end{tabular}
	\end{center}
\end{table}

\subsection{Nuclear Matter Properties and Single-Particle Potentials}
With the couplings fixed in this manner, we first examine the resulting bulk properties of symmetric nuclear matter at saturation density. 
The calculated quantities are listed in Table~\ref{tab:saturation}. 
We obtain a saturation density $\rho_0 = 0.153\,\mathrm{fm^{-3}}$, binding energy per nucleon $E/A = -16.2\,\mathrm{MeV}$, symmetry energy $a_{\rm sym}=27.4\,\mathrm{MeV}$, incompressibility $K=300\,\mathrm{MeV}$, slope parameter $L = 74 \,\mathrm{MeV}$  and effective mass ratio $M^*/M=0.77$. 

The predicted saturation density and binding energy are in good agreement with the accepted empirical values. Moreover, the calculated slope parameter $L$ falls within the empirical range of $58.7\pm28.1$ MeV, while the symmetry energy is somewhat lower than the empirical estimate of $31.7\pm3.2$ MeV, reported in Ref.~\cite{Oertel:2016bki}. On the other hand, the incompressibility exceeds the commonly adopted value of $240\pm20$ MeV summarized by Garg and Col\`o~\cite{Garg:2018uam}. It is important to emphasize that this agreement is nontrivial, as the consistency of the calculated bulk properties indicates that the QCD-derived interaction strengths remain compatible with low energy nuclear phenomenology. In this sense, microscopic constraints imposed at the quark level naturally propagate to realistic bulk nuclear matter behavior within the RMF framework. These bulk nuclear matter properties play a crucial role in determining the stiffness of the hadronic equation of state at supra-nuclear densities and therefore strongly influence the mass–radius relation and the maximum mass of neutron stars.

\begin{table}[H]
	\caption{Properties of the nuclear matter in saturation density.}
	\label{tab:saturation}
	\small
	\begin{center}
		\begin{tabular}{|| c || c | c | c |c | c ||}
			\hline
			$\rho_{0}$ ($fm^{-3}$) & B/A ($MeV$) &  $a_{sym }$ ($MeV$) & K ($MeV$) & L ($MeV$) &$ M^{*}/M$\\ 
			\hline\hline
			
			$0.153$ & $-16.2$ & $27.4$ & $300$ & 74 & $0.77$\\ 
			
			\hline
		\end{tabular}
	\end{center}
\end{table}

The single-particle potentials at saturation density are presented in Table~\ref{tab:potentials}. 
For nucleons we obtain $U_N(\rho_0) \simeq -59\,\mathrm{MeV}$, in agreement with empirical optical-model analyses. 
For hyperons, the calculated potentials are
\begin{equation}
	U_\Lambda(\rho_0) \simeq -30\,\mathrm{MeV}, 
	\qquad
	U_{\Xi^-}(\rho_0) \simeq -10\,\mathrm{MeV},
\end{equation}
which are consistent with constraints extracted from hypernuclear data. 
In particular, the $\Lambda$ potential reproduces the well-established attractive depth of approximately $-28$ to $-30\,\mathrm{MeV}$, while the $\Xi^-$ potential remains moderately attractive and falls within current experimental uncertainties.

\begin{figure}[H]
	\centering
	\includegraphics[width=.48\linewidth]{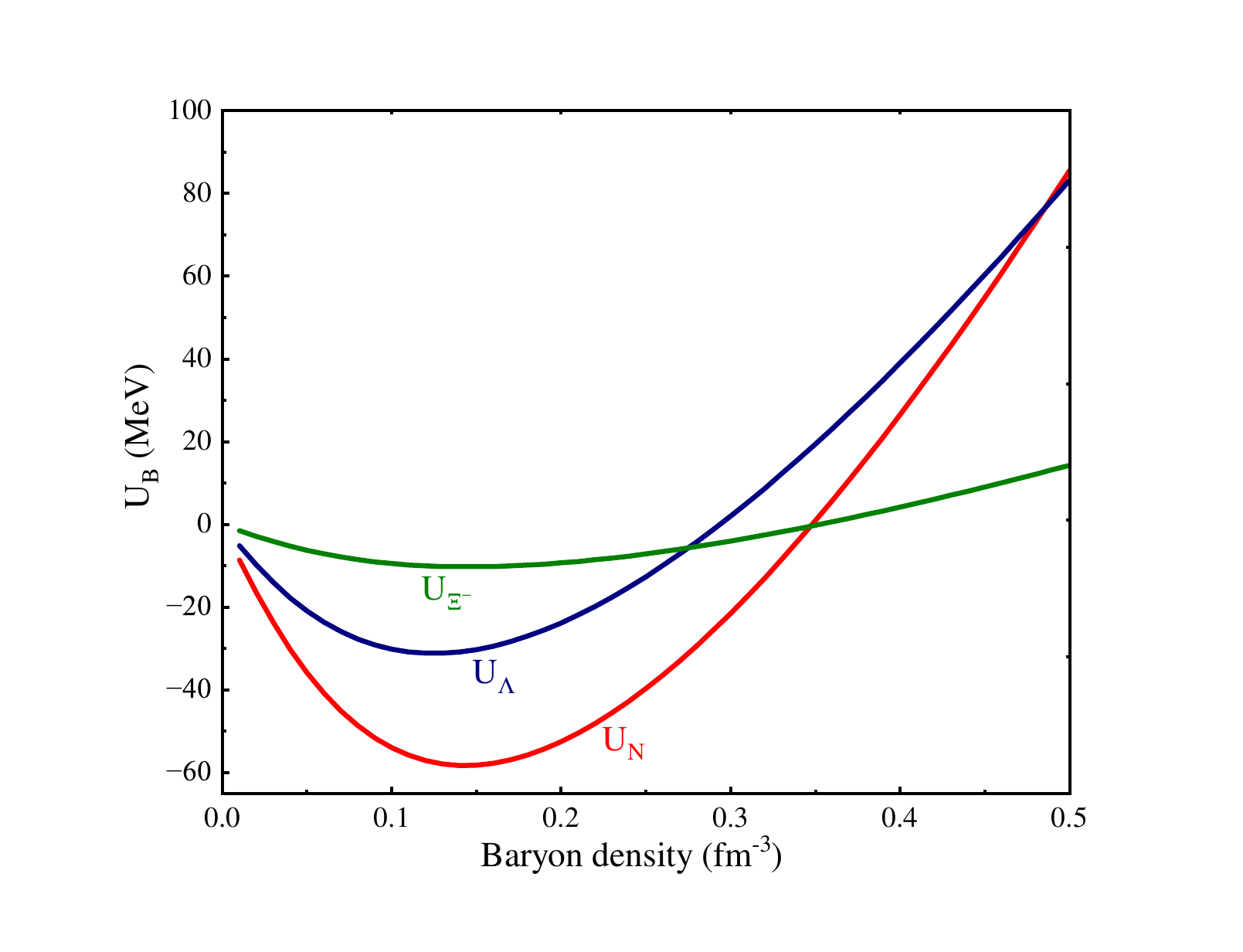}
	\caption{Single potential vs. baryon density.}
	\label{fig:potentials_density}
\end{figure}

The density dependence of the single-particle potentials is displayed in Fig.~\ref{fig:potentials_density}. 
As expected, the nucleon potential becomes increasingly repulsive above saturation density, ensuring sufficient stiffness of the equation of state at supra-nuclear densities. 
In contrast, the hyperon potentials exhibit a smoother density evolution, reflecting the balance between scalar attraction and vector repulsion in the strange sector. 
The baryon single-particle potentials are calculated using a fixed parameter set of coupling constants selected from within the QCD sum-rule allowed intervals. The corresponding mean fields are obtained self-consistently from the RMF field equations. The same parameter set is then employed consistently in the EOS and neutron star calculations, without any further adjustment of the baryonic interaction once the parameter set has been fixed.

Taken together, the results presented in Tables~\ref{tab:saturation}--\ref{tab:potentials} and Fig.~\ref{fig:potentials_density} demonstrate that the hadronic sector of the model simultaneously satisfies 
(i) QCD-based coupling relations, 
(ii) empirical nuclear saturation constraints, and 
(iii) hypernuclear single-particle potential data. 
This multi-level consistency provides a robust and microscopically motivated foundation for extending the equation of state to higher densities and for investigating hyperonic and hybrid star configurations in the following sections.
\begin{table}
	\centering
	\caption{Single-particle potentials at saturation density. 
		Empirical constraints are extracted from analyses of nuclear and hypernuclear data.}
	\label{tab:potentials}
	\begin{tabular}{|| c || c | c ||}
		\hline
		Quantity & This Work (MeV) & Empirical Constraint (MeV) \\
		\hline\hline
		$U_N(\rho_0)$ & $\approx -59$  & $\approx -50$ to $-60$~\cite{SerotWalecka1997}  \\
		$U_\Lambda(\rho_0)$ &  $\approx -30$  & $\approx -28$ to $-30$~\cite{Schaffner_Bielich_2000} \\
		$U_{\Xi^-}(\rho_0)$ &  $\approx -10$ & $\approx -10$ to $-18$~\cite{Schaffner_Bielich_2000} \\
		\hline
	\end{tabular}
\end{table}
As mentioned above, in the present analysis, we restrict the hyperonic sector to the $\Lambda$ and $\Xi^-$ baryons. 
This choice is motivated by the density-dependent composition of charge-neutral $\beta$-equilibrated matter. 
Within such conditions, the $\Lambda$ hyperon typically represents the first strange baryon to appear due to its relatively low threshold under chemical equilibrium. 
The $\Xi^-$ hyperon is also expected to emerge at comparatively low densities, as its negative charge allows it to efficiently reduce the lepton fraction and lower the total free energy of the system.
Since these two species dominate the onset of strangeness in the density regime relevant for neutron star interiors, they capture the leading hyperonic effects on the equation of state. 
For this reason, the present study focuses on the $\Lambda$ and $\Xi^-$ degrees of freedom.

  \subsection{Hybrid star EOS and Particle Fractions}
  
\begin{figure}[H]
  	\centering
  	\subfigure[]{\label{fig:1}
  		\includegraphics*[width=.48\linewidth, height=0.23\paperheight]{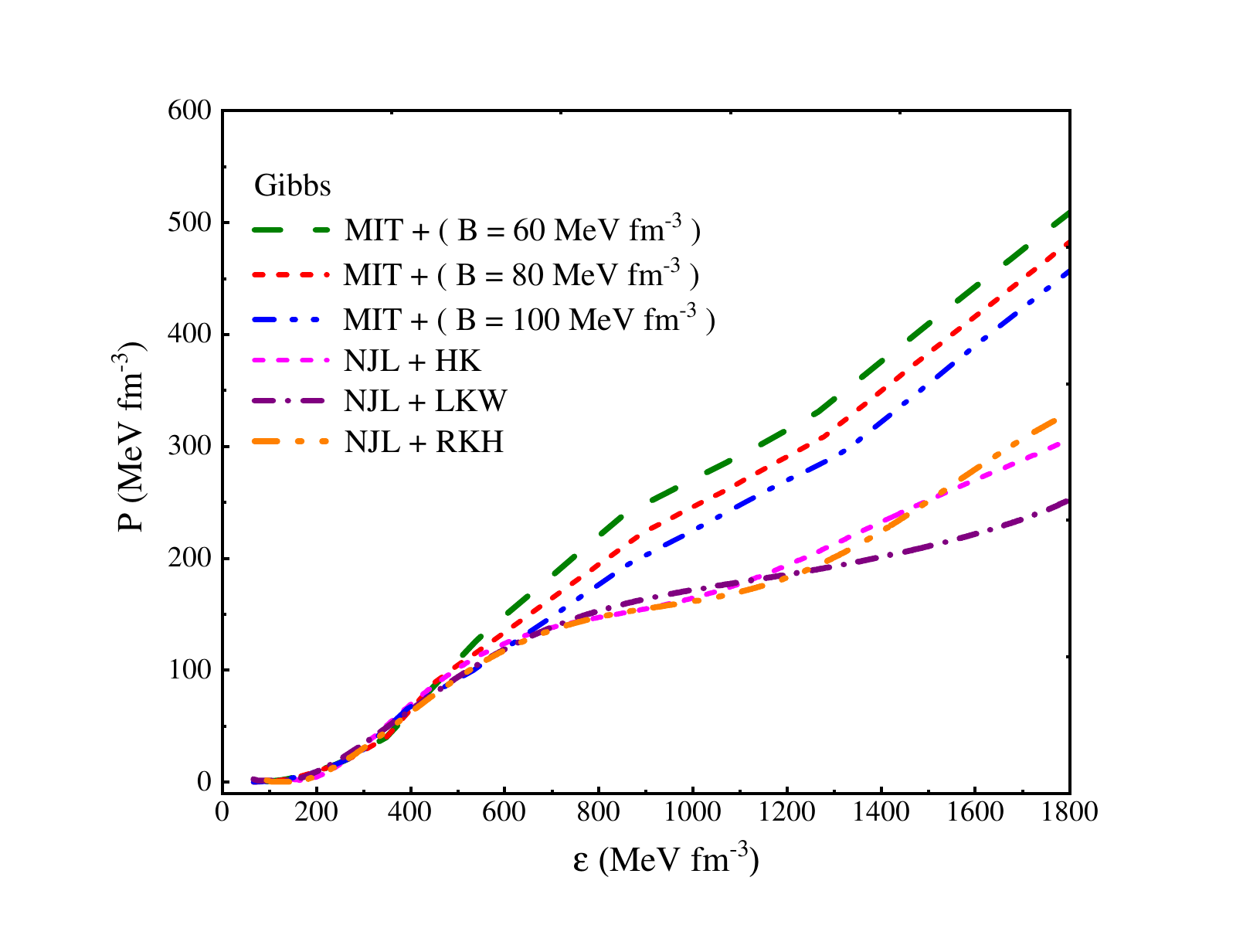}}
  	\hspace{1mm}
  	\subfigure[ ]{\label{fig:2}
  		\includegraphics*[width=.48\linewidth, height=0.23\paperheight]{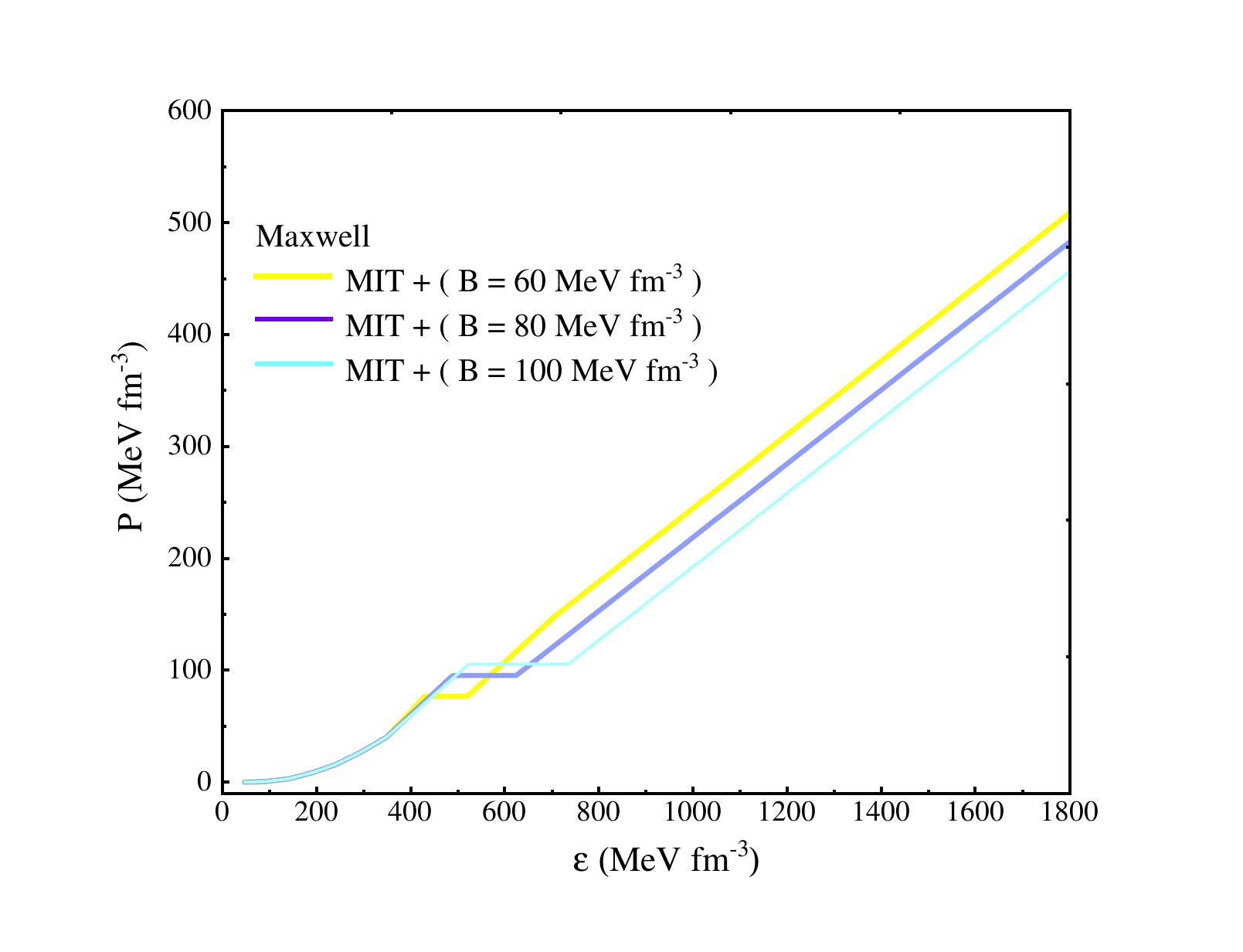}}
  		\caption{Pressure vs. energy density for hybrid matter within the Gibbs ((a)) phase transition combined with the MIT bag model and NJL model, and Maxwell ((b)) phase transition combined with the MIT bag model.}
  	\label{fig:HybridEOS}
  \end{figure}

  Fig.~\ref{fig:HybridEOS} shows the hybrid star matter equations of state obtained using both the Gibbs and Maxwell constructions for the MIT bag model with 
  $B = 60$--$100~\mathrm{MeV\, fm^{-3}}$, together with the results of the NJL model.
  Within the Gibbs construction, the transition from hadronic to quark matter occurs smoothly over a finite density interval. Consequently, the pressure increases continuously with energy density and no plateau structure appears in the EOS. The mixed phase is characterized by the gradual conversion of hadronic matter into quark matter under the condition of global charge neutrality.

  For the MIT model, increasing the bag constant from 
  $60$ to $100~\mathrm{MeV\, fm^{-3}}$ systematically lowers the pressure at a given energy density. In particular, the EOS corresponding to 
  $B = 60~\mathrm{MeV\, fm^{-3}}$ lies above the other curves and exhibits the stiffest behavior, whereas 
  $B = 100~\mathrm{MeV\, fm^{-3}}$ produces the softest EOS in the high-density region. Therefore, within the present parameter set, a larger bag constant results in a softer hybrid equation of state. This behavior can be understood from the structure of the MIT model, where the bag constant represents the vacuum energy difference between the confined and deconfined phases. A larger $B$ increases the vacuum energy cost of quark matter formation, effectively reducing the quark pressure and softening the EOS.
  In comparison, the NJL model leads to a distinct behavior in the transition region due to its dynamical treatment of quark interactions.
  
\begin{figure}[H]
	\centering
	\subfigure[]{\label{fig:1}
		\includegraphics*[width=.35\linewidth, height=0.19\paperheight]{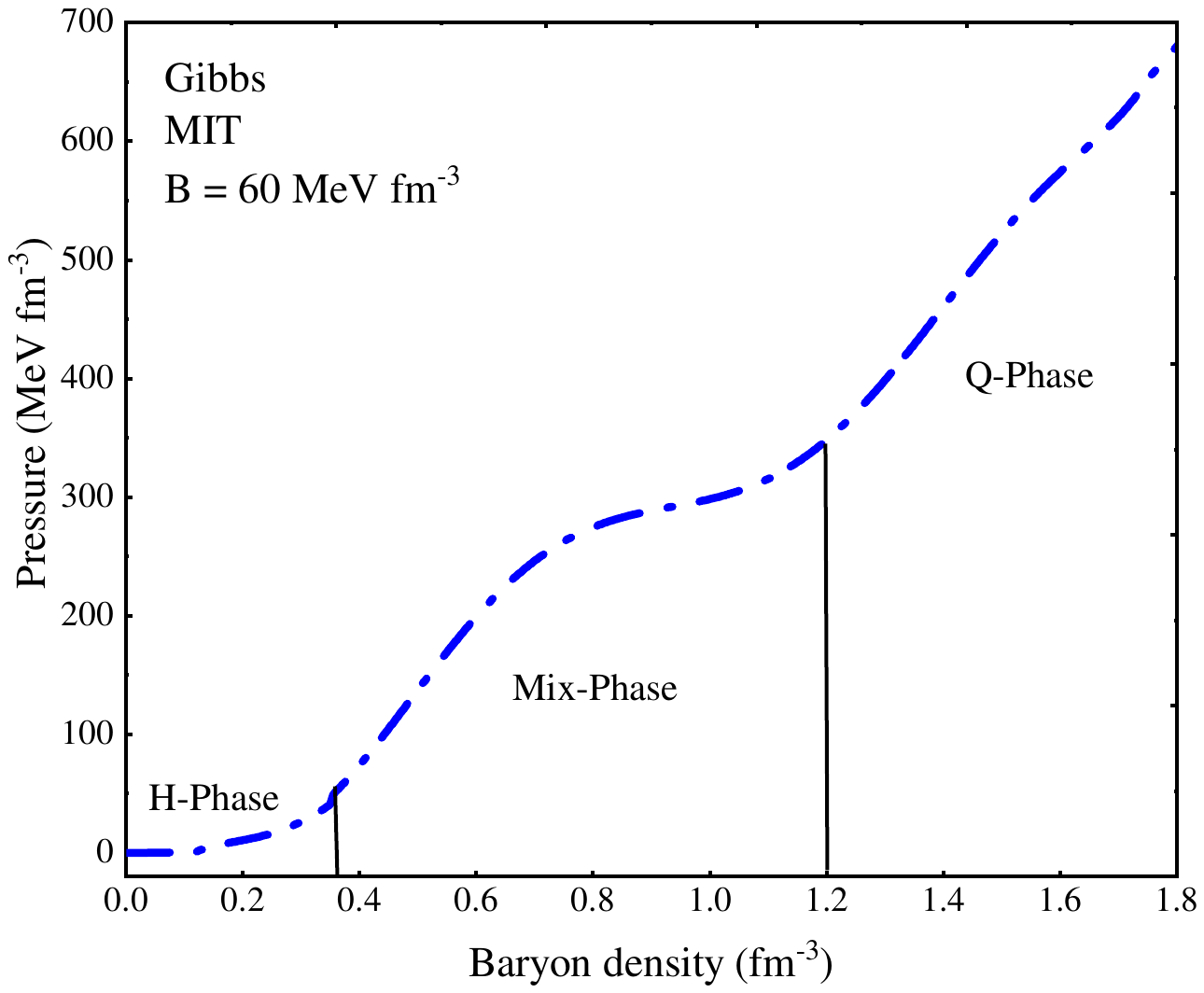}}
	\hspace{1mm}
	\subfigure[ ]{\label{fig:2}
		\includegraphics*[width=.35\linewidth, height=0.19\paperheight]{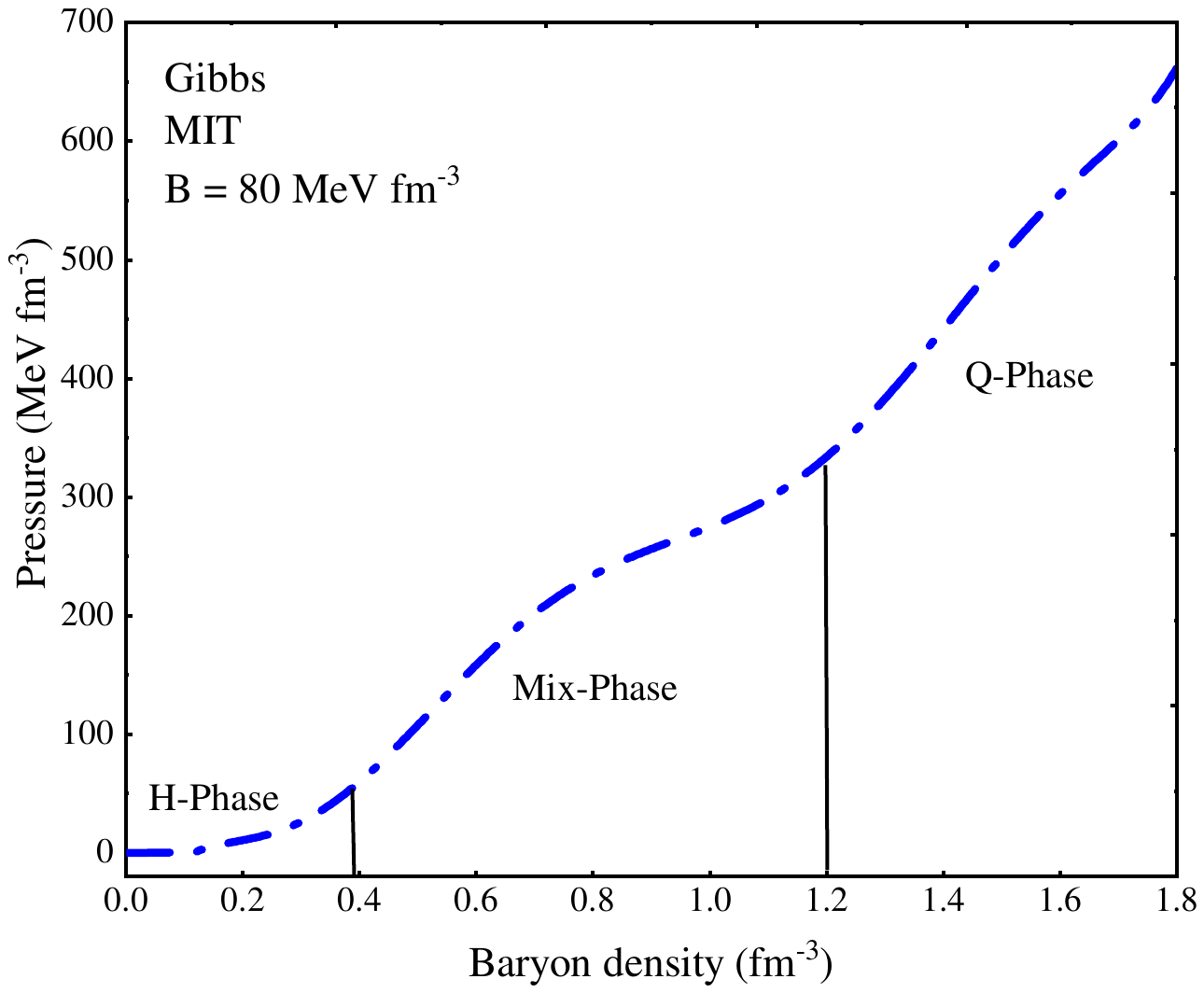}}
	\hspace{1mm}
	\subfigure[ ]{\label{fig:3}
		\includegraphics*[width=.35\linewidth]{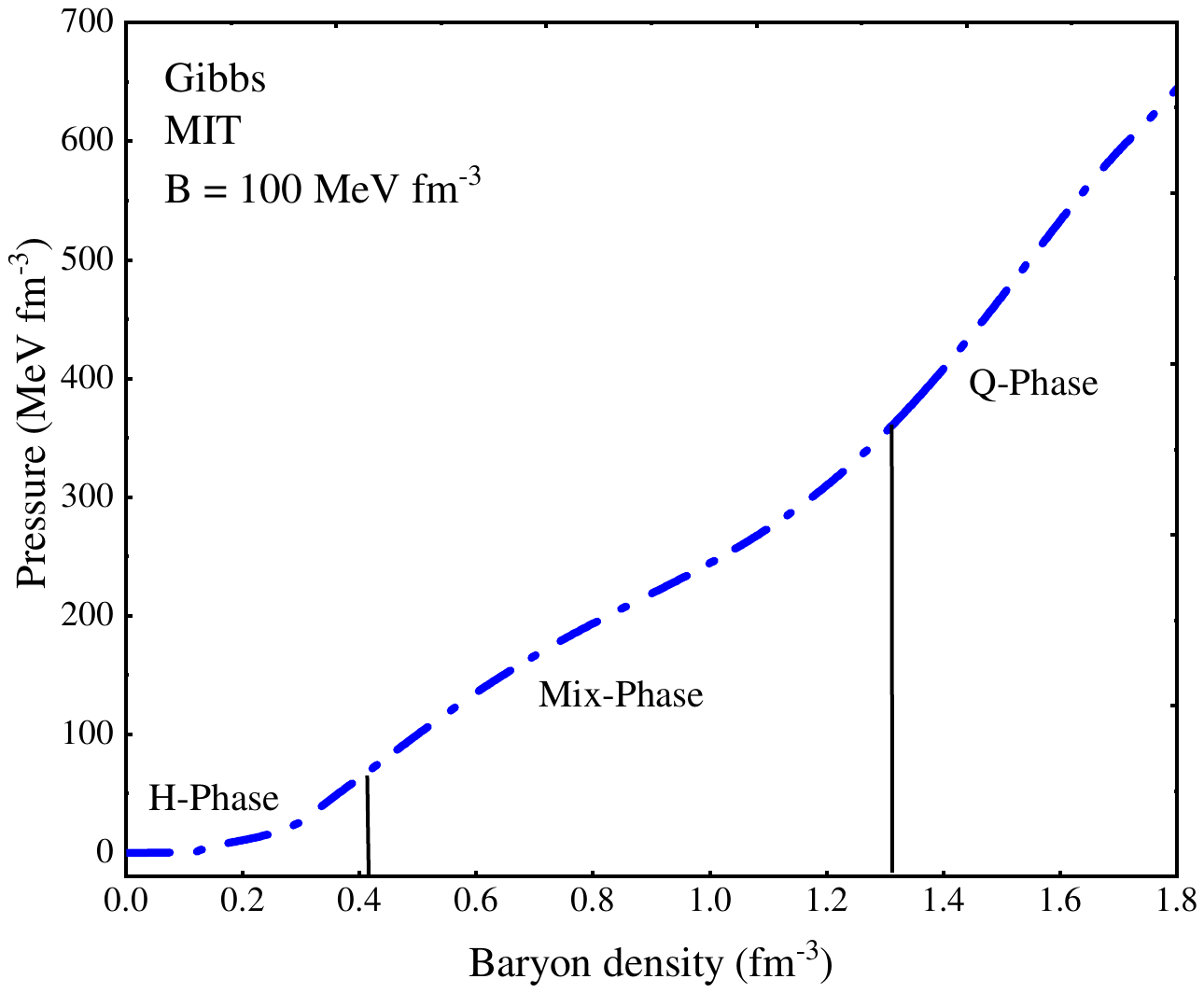}}
	\hspace{1mm}
	\subfigure[ ]{\label{fig:4}
		\includegraphics*[width=0.35\linewidth]{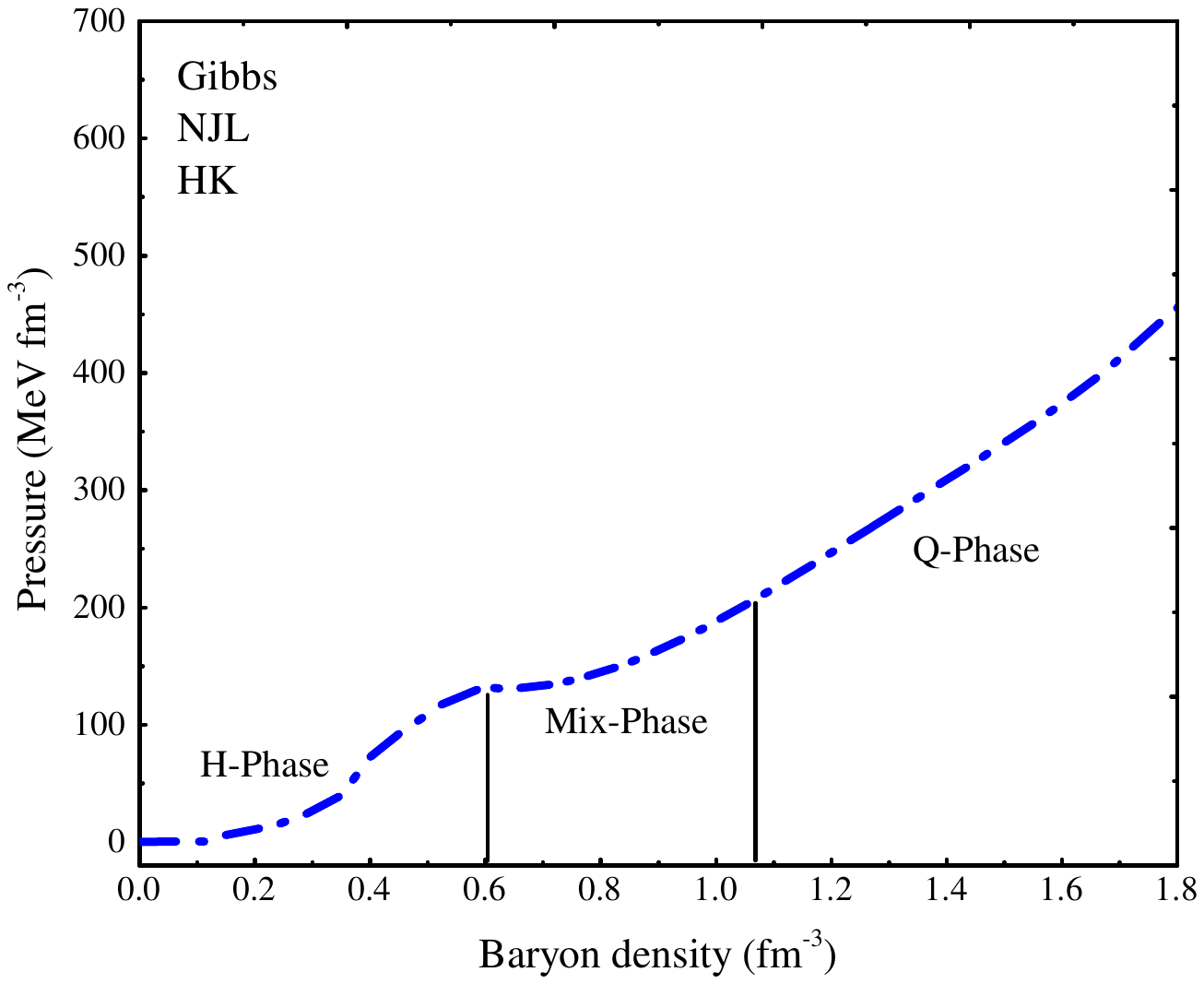}}
			\hspace{1mm}
		\subfigure[ ]{\label{fig:5}
			\includegraphics*[width=.35\linewidth]{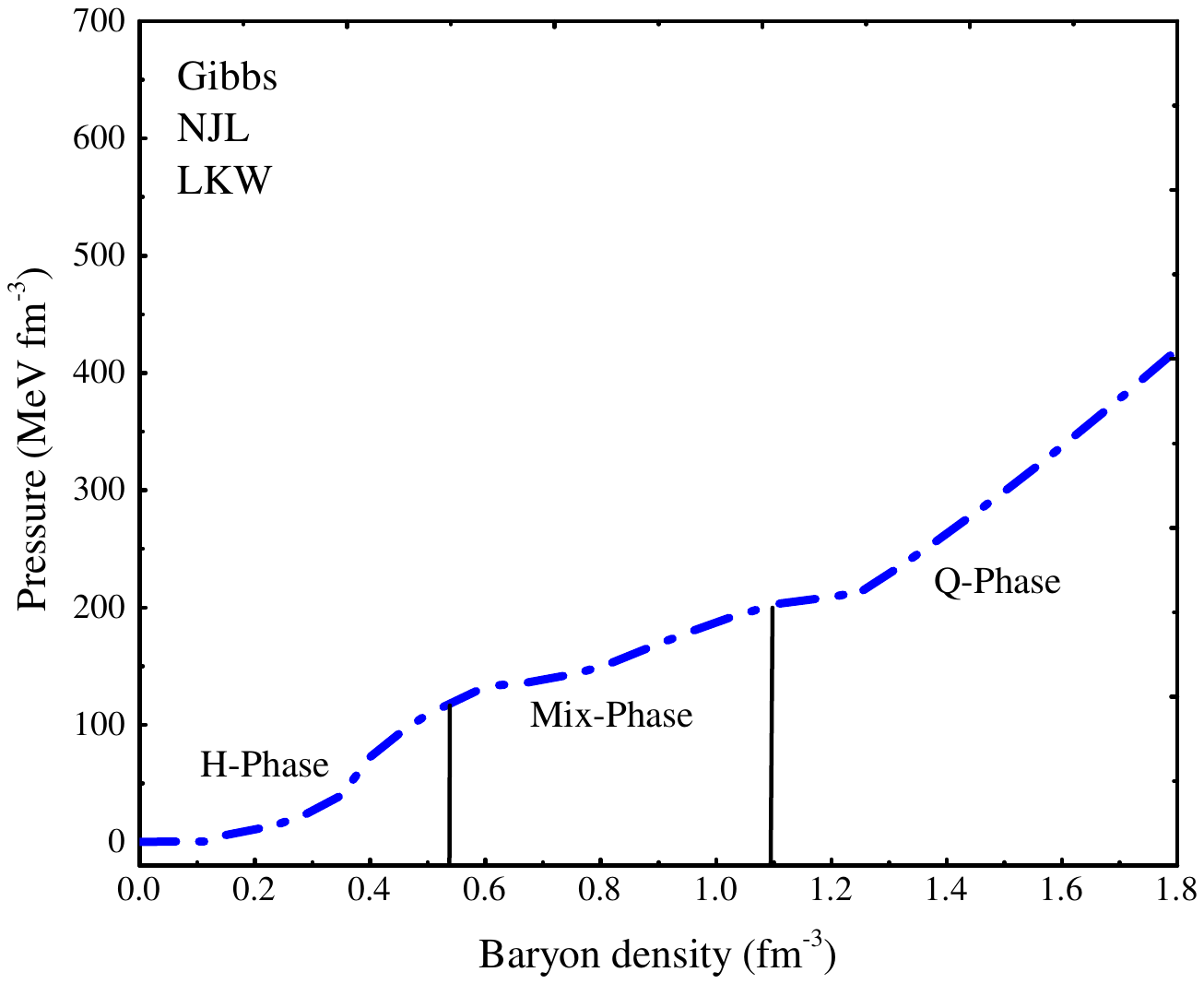}}
		\hspace{1mm}
			\hspace{1mm}
		\subfigure[ ]{\label{fig:6}
			\includegraphics*[width=.35\linewidth]{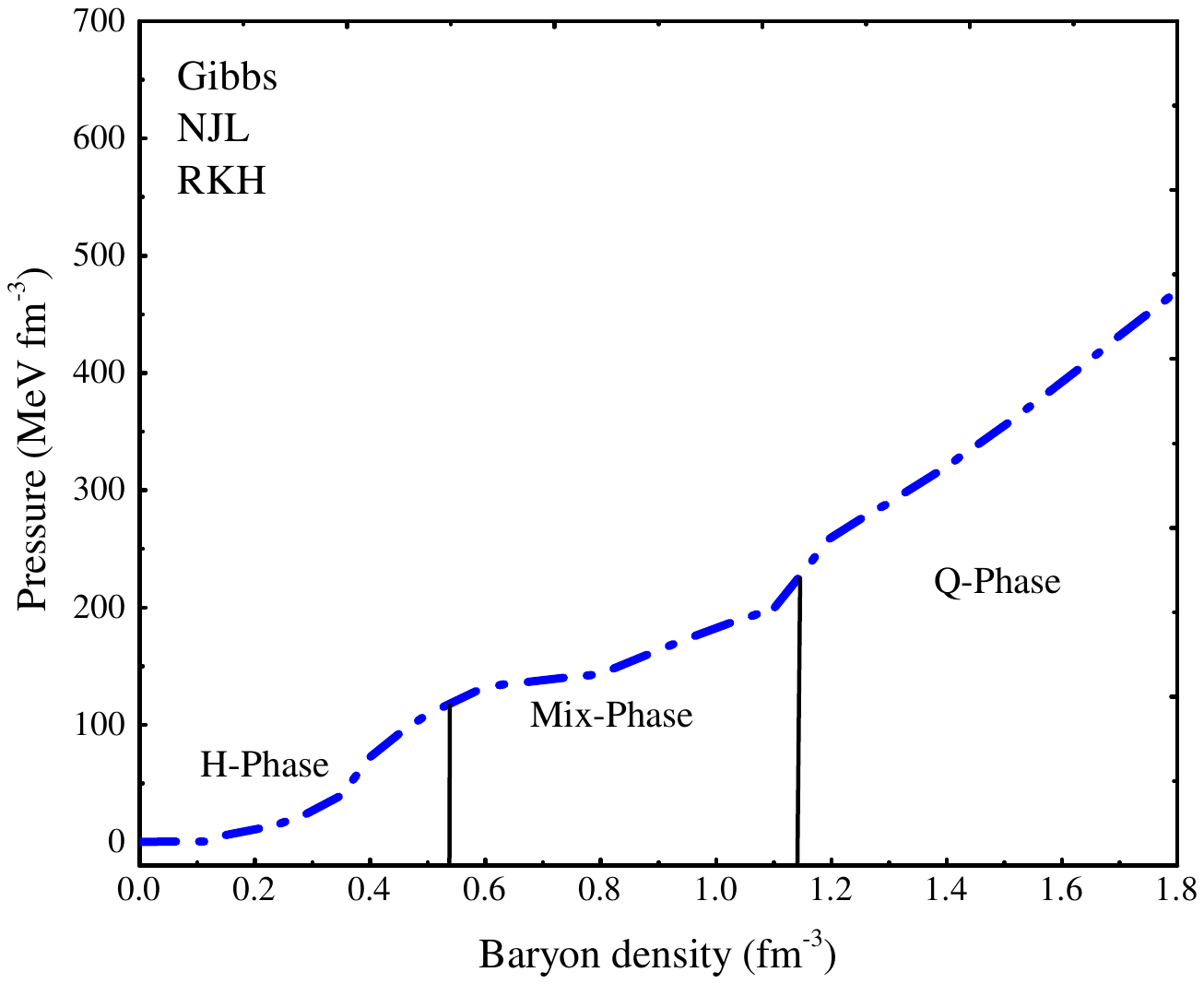}}
				\hspace{1mm}
			\subfigure[ ]{\label{fig:7}
				\includegraphics*[width=.35\linewidth]{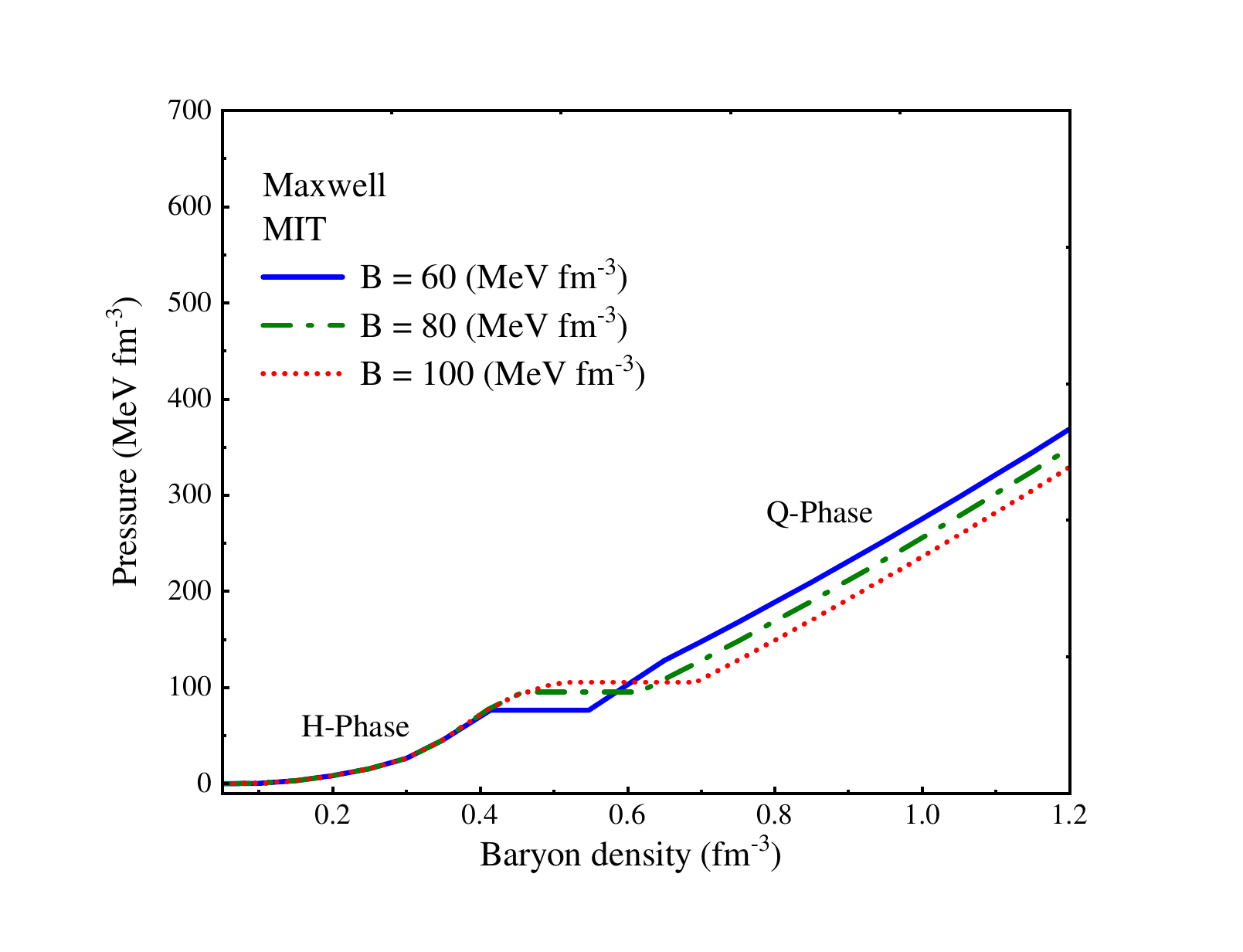}}

	\caption{Pressure vs. baryon density for hybrid matter within the
		Gibbs (combined with the MIT bag model and NJL model) and Maxwell (combined with the MIT bag model) phase transition.}
	\label{fig:PhaseDensity}
\end{figure}

The absence of an explicit confinement term modifies the high-density stiffness of quark matter and consequently affects the overall structure of the hybrid EOS.
The phase structure of hybrid matter is further illustrated in Fig.~\ref{fig:PhaseDensity}, where the pressure density is plotted as a function of baryon density for both Gibbs and Maxwell constructions. In the Gibbs scenario, three distinct regions can be identified: the purely hadronic phase at low densities (from: 0 to 0.37 $\,\mathrm{fm^{-3}}$ , 0 to 0.39 $\,\mathrm{fm^{-3}}$, 0 to 0.41 $\,\mathrm{fm^{-3}}$ for $B=60, 80, 100~\mathrm{MeV\, fm^{-3}}$ respectively), the mixed phase over a finite density interval (from: 0.37 to 1.2 $\,\mathrm{fm^{-3}}$, 0.39 to 1.2 $\,\mathrm{fm^{-3}}$, 0.41 to 1.3 $\,\mathrm{fm^{-3}}$ for $B=60, 80, 100~\mathrm{MeV\, fm^{-3}}$ respectively), and the pure quark phase at high densities. The transition is continuous, and both the pressure density and baryon density vary smoothly across the mixed phase. No discontinuity is observed, reflecting the global charge neutrality condition and the coexistence of hadronic and quark components. In contrast, the Maxwell construction leads to a sharp first-order phase transition. At the transition pressure, the system undergoes a discontinuous jump in baryon density (from: 0.40 to 0.54 $\,\mathrm{fm^{-3}}$, 0.46 to 0.62 $\,\mathrm{fm^{-3}}$, 0.52 to 0.69 $\,\mathrm{fm^{-3}}$ for $B=60, 80, 100~\mathrm{MeV\, fm^{-3}}$ respectively) while the pressure remains constant. Consequently, no extended mixed phase region appears in this case, and matter converts abruptly from a purely hadronic phase to a pure quark phase. It is important to emphasize that for the NJL model the Maxwell construction cannot be realized within the present parameter set. The quark pressure remains below the hadronic pressure for all relevant baryon chemical potentials, and therefore the condition $P_H(\mu_B) = P_Q(\mu_B)$ is never satisfied. As a result, a Maxwell-type first-order phase transition does not occur for the NJL model. This behavior originates from the relatively soft quark pressure predicted by the NJL model within the present parameter sets, which prevents the quark EOS from overtaking the hadronic EOS in the relevant chemical‑potential range. The appearance of hyperons partially softens the hadronic EOS at intermediate densities, which can influence the onset density of quark matter and modify the width of the mixed phase. Also, in the Maxwell construction, each phase is required to satisfy local charge neutrality independently, which prevents the formation of a spatially extended mixed phase.

\begin{figure}[H]
	\centering
	\subfigure[]{\label{fig:41a}
		\includegraphics*[width=.48\linewidth, height=0.23\paperheight]{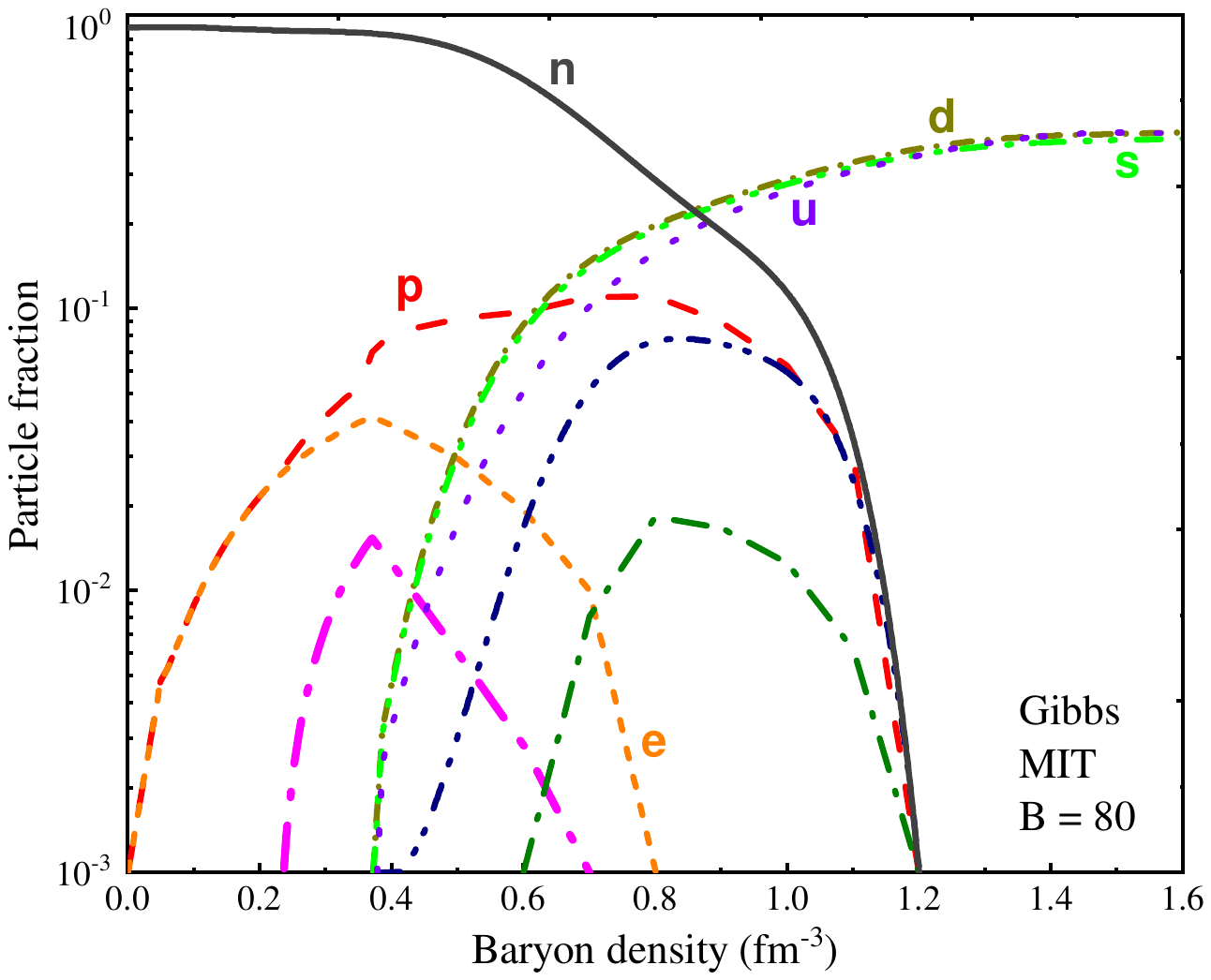}}
	\hspace{1mm}
	\subfigure[ ]{\label{fig:42b}
		\includegraphics*[width=.48\linewidth]{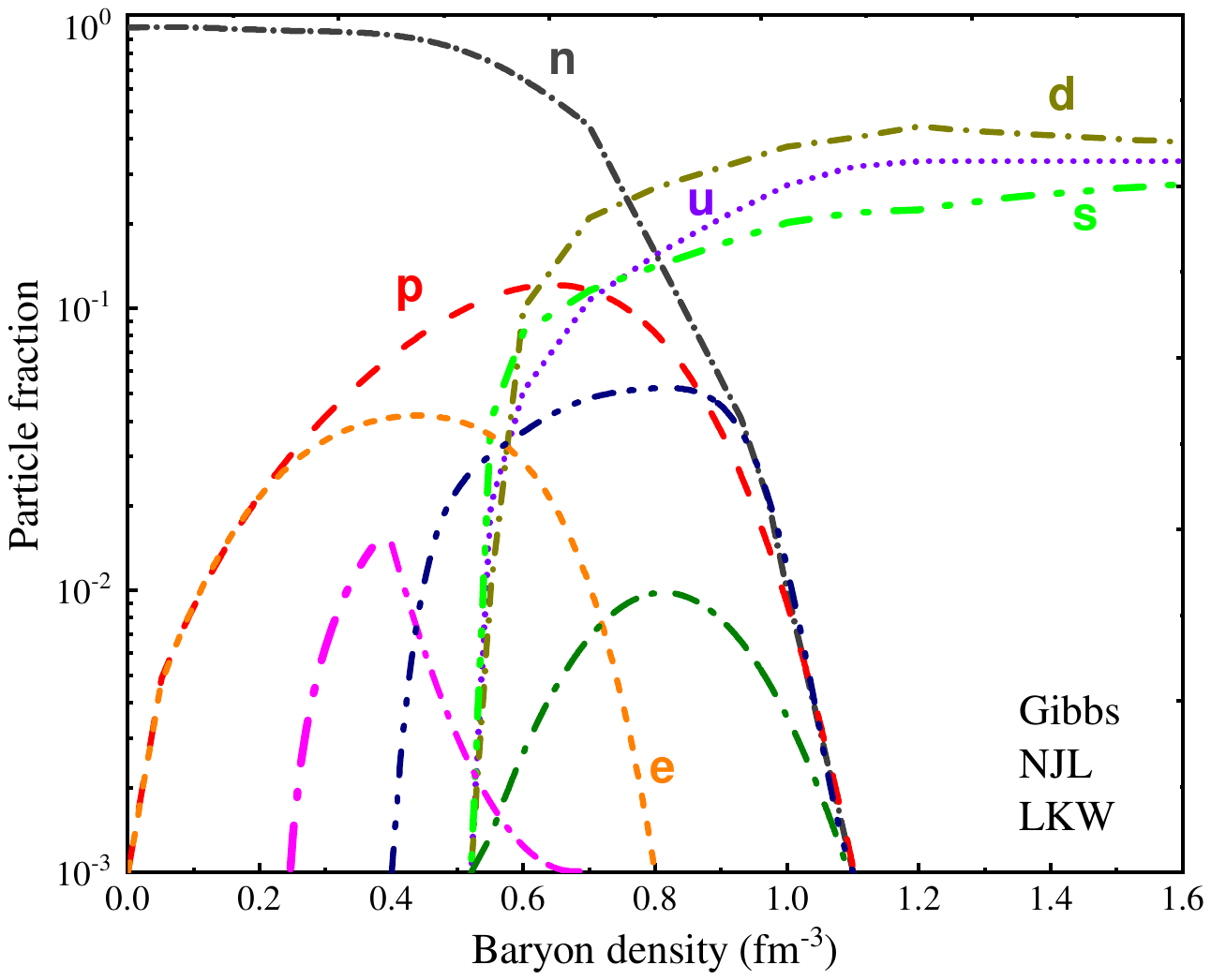}}
		
			\hspace{1mm}
		\subfigure[ ]{\label{fig:43c}
			\includegraphics*[width=.48\linewidth]{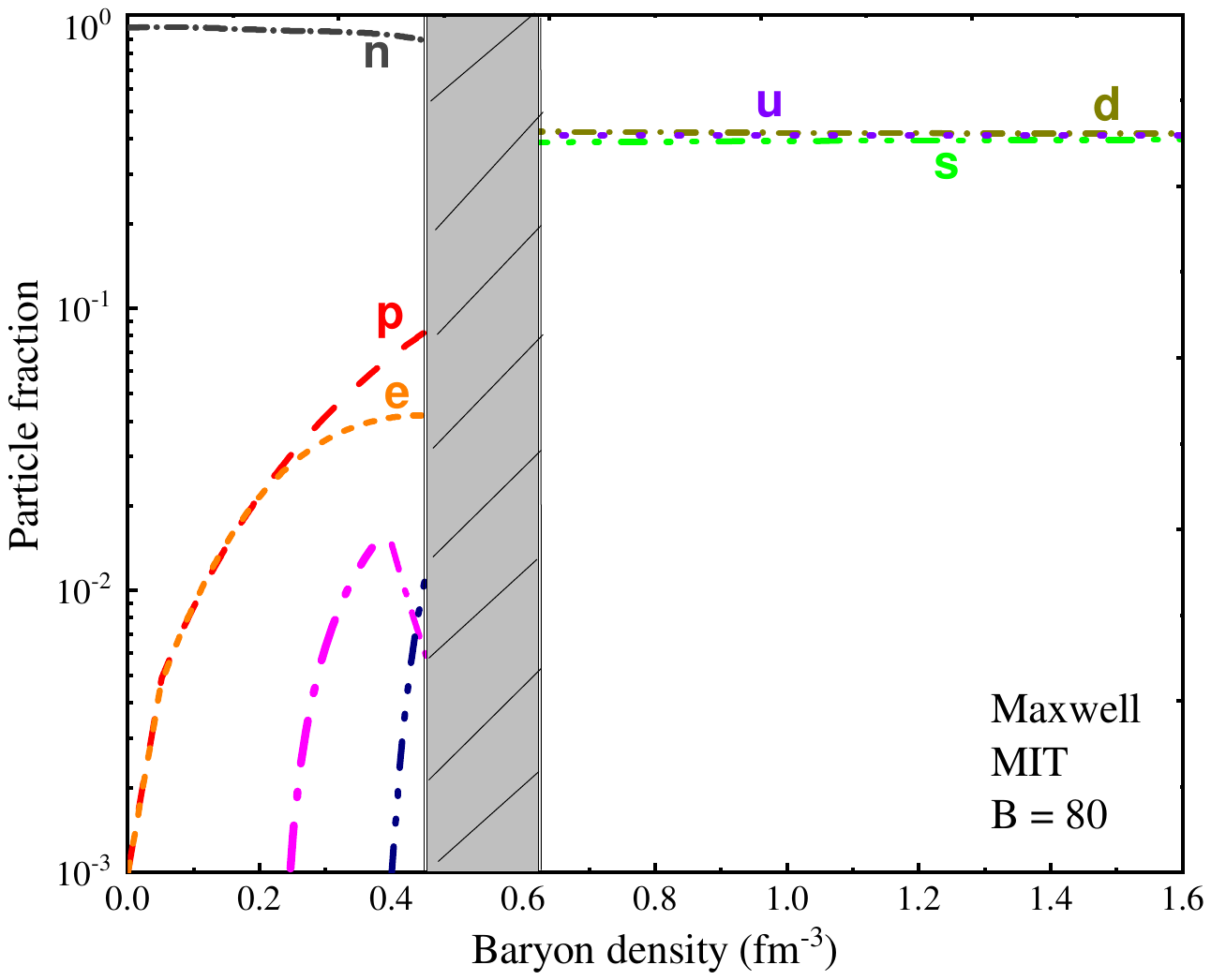}}
	\caption{Particle fraction for hybrid matter within the
		Gibbs and Maxwell phase transition combined with the MIT bag model and NJL model. (a): the Gibbs construction combined with the MIT bag model for $80~\mathrm{MeV\, fm^{-3}}$, (b): the Gibbs construction with the NJL (LKW) model, and (c): the Maxwell construction with the MIT model for $80~\mathrm{MeV\, fm^{-3}}$.}
	\label{fig:ParticleFractions}
\end{figure}

 The evolution of particle fractions as a function of baryon density is shown in Fig.~\ref{fig:ParticleFractions}, providing a detailed picture of the internal composition of matter and clearly illustrating the qualitative differences between the Gibbs and Maxwell constructions.
 Within the Gibbs construction, the quark volume fraction increases continuously from zero to around unity throughout the mixed phase region. This behavior reflects the gradual replacement of hadronic degrees of freedom by deconfined quarks. In this region, the matter composition may simultaneously include nucleons, leptons, quarks, and, depending on the parameter set, hyperons. The relative fraction of each species depends on the density and the specific choice of model parameters. As the density increases, quarks progressively dominate the composition in the high-density regime.
 In contrast, the Maxwell construction exhibits an abrupt change in particle composition. The system transitions directly from a purely hadronic phase (including nucleons and, if present, hyperons) to a pure quark phase, without an extended coexistence region. Consequently, the particle fractions show a discontinuous behavior consistent with a sharp first-order phase transition.
 Overall, the microscopic composition of matter strongly depends on both the chosen quark model and the phase transition construction. These differences in internal structure are expected to have direct implications for the macroscopic properties of compact stars, particularly their mass–radius relation and central composition.
 For clarity and to avoid unnecessary repetition, we do not display all nine computed particle fraction in stars. Instead, three representative cases are presented, which capture the essential qualitative behavior of the different phase-transition scenarios and quark matter models. Specifically, we show: the Gibbs construction combined with the MIT bag model for $80~\mathrm{MeV\, fm^{-3}}$ (Fig. \ref{fig:41a}), the Gibbs construction with the NJL (RKH) model (Fig. \ref{fig:42b}), and the Maxwell construction with the MIT model for $80~\mathrm{MeV\, fm^{-3}}$ (Fig. \ref{fig:43c}). The remaining configurations exhibit the same qualitative trends and differ only quantitatively. Therefore, their omission from the figures does not affect the physical interpretation or the conclusions drawn from the analysis.
 Figures 4(a–c) illustrate the evolution of particle fractions in hybrid matter under different phase transition constructions and microphysical models. In Fig. \ref{fig:41a}, corresponding to the Gibbs construction within the MIT bag model ($B = 80~\mathrm{MeV\, fm^{-3}}$), the mixed phase begins at a baryon density of approximately 0.39  $\,\mathrm{fm^{-3}}$, signaling the onset of quark deconfinement. The fractions of u, d, and s-quarks increase steadily throughout the coexistence region, marking the gradual formation of quark matter. Hyperons such as $\Lambda$ and $\Xi^{-}$ appear at densities around 0.41 $\,\mathrm{fm^{-3}}$ and 0.6 $\,\mathrm{fm^{-3}}$, respectively, indicating the role of strangeness at intermediate densities. As the system approaches the upper boundary of the mixed phase, near 1.2 $\,\mathrm{fm^{-3}}$, baryonic and hyperonic contributions diminish to nearly zero, while quark fractions dominate the matter composition toward the stellar core. This behavior demonstrates a smooth and thermodynamically consistent transition between hadronic and quark matter in the Gibbs approach.
 In Fig. \ref{fig:42b}, obtained using the NJL–LKW model under the Gibbs formalism, the qualitative trend remains similar but with subtle physical differences. Notably, the $\Lambda$ hyperon appears prior to the onset of the mixed phase, suggesting an earlier participation of strange baryons in the hadronic sector. The subsequent rise of quark fractions (u, d, s) beyond the transition density indicates a gentler conversion toward quark matter compared to the MIT case. The coexistence region extends up to roughly 1.2  $\,\mathrm{fm^{-3}}$, where the matter becomes completely deconfined. The shift in the onset density and the evolution of strangeness highlight the sensitivity of hybrid matter composition to the underlying quark interaction model.  
 In Fig. \ref{fig:43c}, the Maxwell construction with the MIT bag model is applied. Here, the phase transition occurs discontinuously at a critical baryon density of about 0.46  $\,\mathrm{fm^{-3}}$, with no intermediate mixed region. The hadronic components vanish abruptly at the transition point, and the quark fractions instantly dominate beyond it. This sharp boundary reflects the first–order nature of the Maxwell phase transition, at which the hadronic and quark phases have the same transition pressure and baryon chemical potential, while the baryon density changes discontinuously across the interface. In summary, while both the Gibbs and Maxwell approaches describe the same fundamental hadron–quark transition, their physical implications differ substantially. The Gibbs construction produces a smoother transition with coexistence of hadronic and quark degrees of freedom, whereas the Maxwell construction yields an abrupt, first-order transition at the critical density. Moreover, the choice between the MIT and NJL–LKW models influences the onset of strangeness and the internal composition of hybrid stars, ultimately affecting macroscopic properties such as maximum mass, radius, and the stability of compact star configurations.

\subsection{Hybrid star structure}
	\label{TOV}
The Tolman-Oppenheimer-Volkoff (TOV) equations \cite{Tolman:1939jz, Oppenheimer:1939ne} describe the equilibrium structure of a spherically symmetric compact star in general relativity. They relate the radial pressure gradient to the local energy density and enclosed gravitational mass. The equations (in natural unit) are written as:
\begin{equation}
	\label{TOV1}
	\dfrac{dP(r)}{dr} = - \dfrac{M(r) \varepsilon(r)}{r^{2}}(1+ \dfrac{P(r)}{\varepsilon(r)})(1+\dfrac{4\pi r^{3} P(r)}{M(r)}) (1- \dfrac{2M(r)}{r})^{-1},
\end{equation}
and
\begin{equation}
	\dfrac{dM(r)}{dr} = {4\pi\varepsilon(r)r^{2}}.
\end{equation}
Here $P(r)$, $\varepsilon(r)$, and $M(r)$ denote the pressure, energy density, and enclosed mass at radius $r$. To obtain stellar configurations, the system is integrated outward from an assumed central density, $\rho_{C}$, until the pressure vanishes, $P(R)=0$, which defines the stellar radius $R$. The gravitational mass then follows as   
\begin{equation}
	M(R^{'}) =  \int_0^{R^{'}} 4\pi r^{2} \varepsilon(r)dr.
\end{equation} 
The solutions require an equation of state (EOS) relating pressure and energy density. In this work, the hybrid EOS models described in Sec. \ref{sec:two} are employed for the stellar interior, while the BPS and BBP EOSs \cite{PPS, Baym:1971ax} are used to describe the outer and inner crust, respectively. The TOV equations are solved numerically using a Runge–Kutta method. Repeating the integration for a range of central densities yields a sequence of equilibrium configurations from which the mass–radius (M–R) relation and maximum mass are determined.

\begin{figure}[H]
	\centering
	\subfigure[]{\label{fig:5a}
		\includegraphics*[width=.48\linewidth, height=0.23\paperheight]{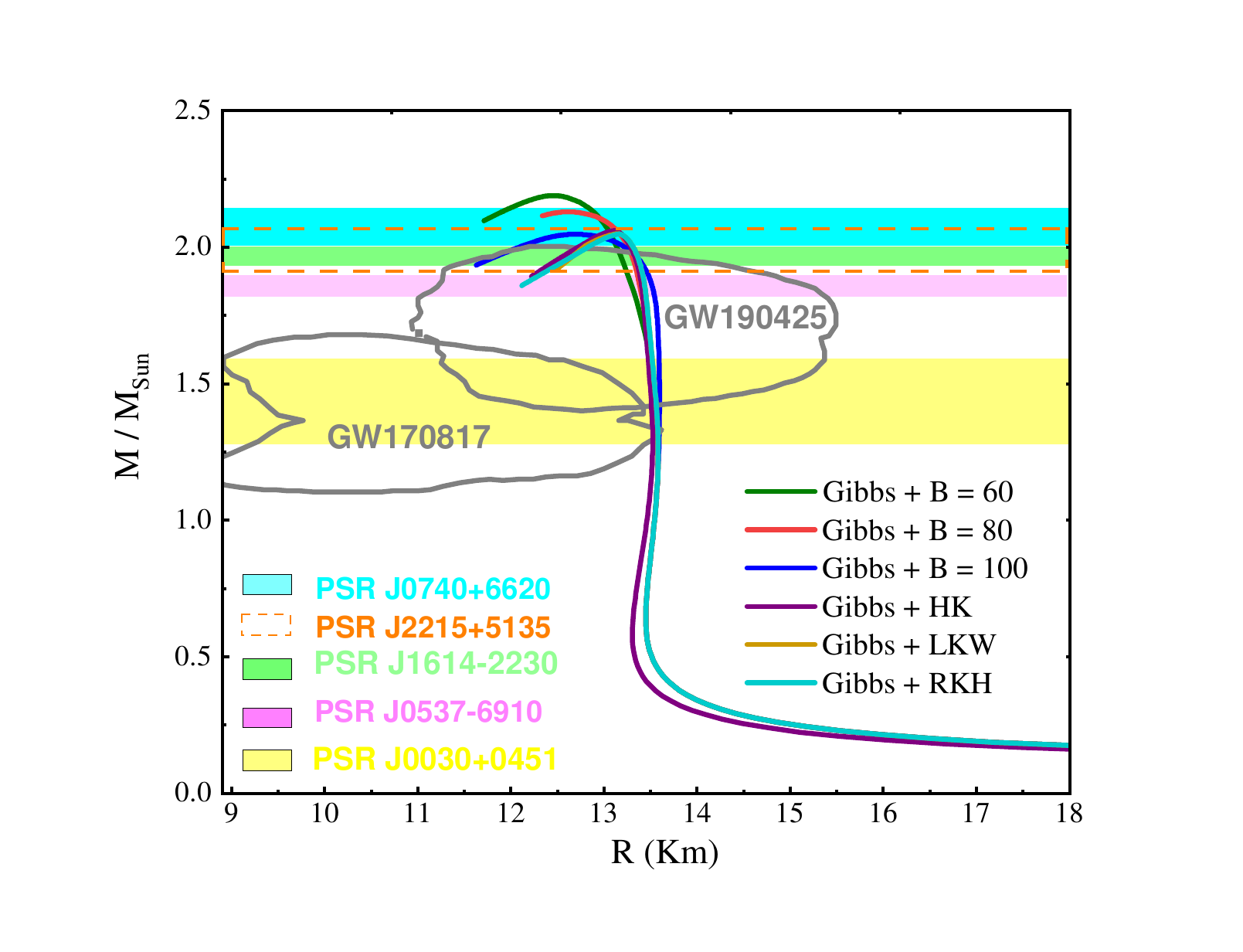}}
	\hspace{1mm}
	\subfigure[ ]{\label{fig:5b}
		\includegraphics*[width=.48\linewidth, height=0.23\paperheight]{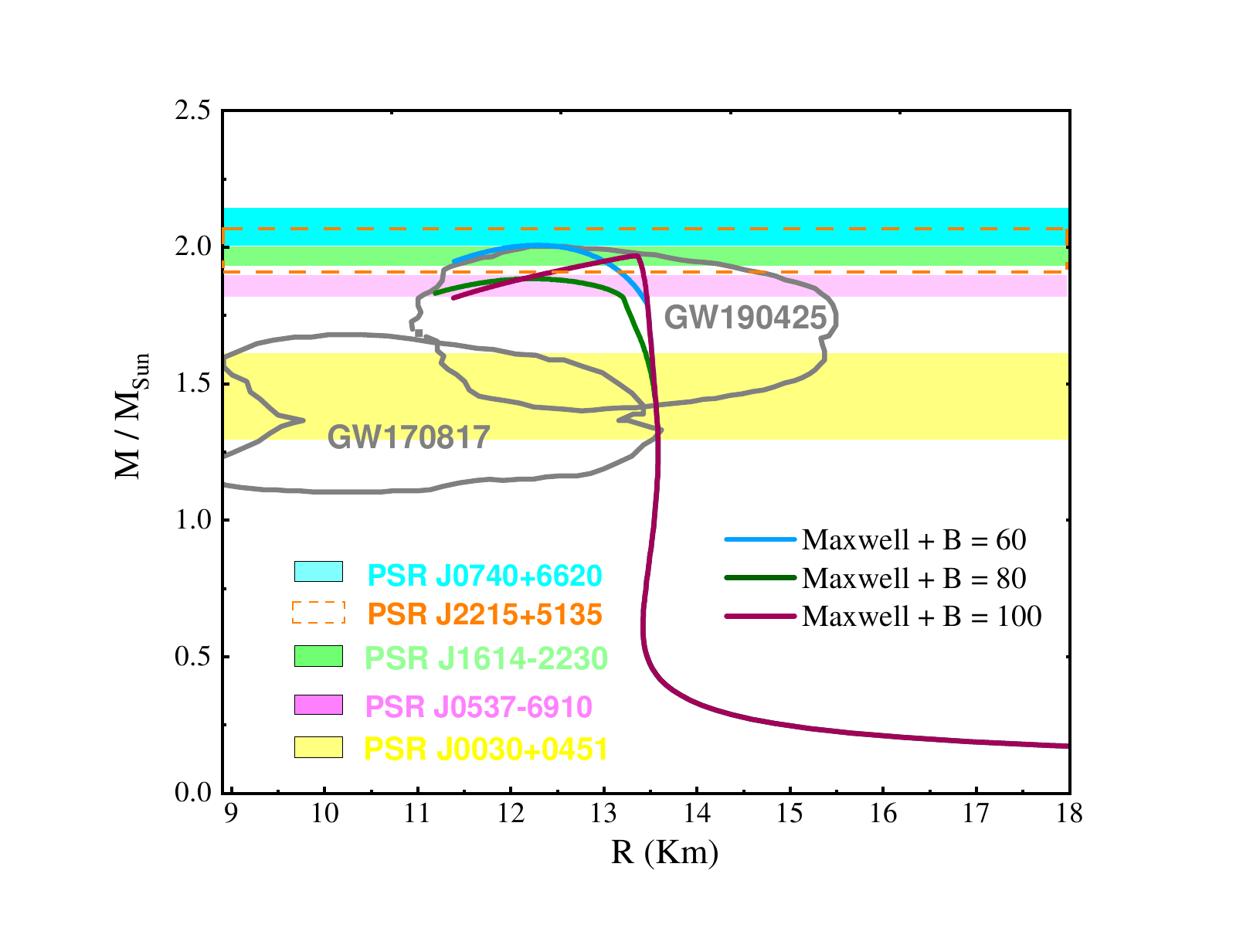}}
	\caption{The mass-radius relation for hybrid matter within the Gibbs ((a)) phase transition combined with the MIT bag model and NJL model, and Maxwell ((b)) phase transition combined with the MIT bag model in compared with observational data.}
	\label{fig:r-m-tot}
\end{figure}

Fig.~\ref{fig:r-m-tot} displays the resulting mass–radius curves for all EOSs considered, together with current observational constraints from massive pulsars and gravitational-wave events. Any viable EOS must support neutron stars with masses of at least two solar masses or higher. A systematic comparison between the Gibbs and Maxwell constructions reveals a clear and robust trend. all Gibbs constructions—whether based on the MIT bag model (for $B=60, 80, 100~\mathrm{MeV\, fm^{-3}}$) or on the NJL parametrizations (RKH, HK, LKW)—predict maximum masses above the two-solar-mass threshold. This reflects the different density dependence of the pressure through the mixed-phase region. In contrast, the Maxwell construction exhibits a more restrictive behavior. Only the Maxwell + MIT configuration with $B=60~\mathrm{MeV\, fm^{-3}}$ is able to generate stellar configurations with $M_{\rm max} = 2\,M_{\odot}$. For larger bag constants, such as $B=80~\mathrm{MeV\, fm^{-3}}$, the predicted maximum mass falls below the two-solar-mass limit. This reduction originates from the softening of the equation of state together with the sharp density discontinuity at the phase boundary, which accelerates the onset of gravitational instability.

The numerical values summarized in Table~\ref{table:m-r} confirm these findings. The Gibbs + MIT ($B=60$) model yields the largest maximum mass, $M_{\rm max}=2.18\,M_{\odot}$, with a corresponding radius of $12.45\,\mathrm{km}$. The Gibbs + MIT ($B=100$) and Gibbs + NJL configurations also satisfy the observational lower bound of two solar masses, albeit with slightly smaller maximum masses. In contrast, the Maxwell + MIT ($B=80$) case predicts $M_{\rm max}=1.88\,M_{\odot}$. While this remains compatible with several observationally inferred neutron star masses, it falls short of reaching the heaviest observed stars exceeding two solar masses.

\begin{table}[H]
	\caption{The maximum masses ($M_{max}/M_{\odot}$) and radii ($ R $) values of stellar matter for each set and observed object.}
	\label{table:m-r}
	\small
	\begin{center}
		\begin{tabular}{|| c || c | c | c ||}
			\hline
			The EOS &$\rho_{central}$ $(fm^{-3})$ & $M_{max}/M_{\odot}$ & Radius ($Km$)  \\ 
			\hline\hline
			\multicolumn{4}{||c||}{Gibbs phase transition + MIT bag model} \\
			\cline{1-4}
			B = 60 & $0.84$ & $2.18$ & $12.45$  \\ 
			B = 80 & $0.76$ & $2.12$ & $12.61$  \\
			B = 100 & $0.77$ & $2.05$ & $12.70$  \\
			\hline\hline
			\multicolumn{4}{||c||}{Gibbs phase transition + NJL model} \\
			\cline{1-4}
			RKH & $0.79 $ & $2.05$ & $13.16 $  \\ 
			HK & $0.80$ & $2.06 $ & $13.11$  \\ 
			LKW & $0.77$ & $2.05$ & $13.14$  \\ 
			\hline\hline
			\multicolumn{4}{||c||}{Maxwell phase transition + MIT bag model} \\
			\cline{1-4}
			B = 60 & $0.95$ & $2.00$ & $12.28$  \\ 
			B = 80 & $0.93$ & $1.88$ & $12.22$  \\
			B = 100 & $0.70$ & $1.96$ & $13.33$  \\
			\hline\hline
			\multicolumn{4}{||c||}{Observed Data} \\
			\cline{1-4}
			PSR~J0740+6620 \cite{Fonseca:2021wxt} & - & $2.07 \pm 0.07 $& $12.39^{+1.30}_{-0.98}$  \\ 
				PSR~J2215+5135 \cite{Sullivan:2024qgl} & - & $1.98 \pm 0.08 $& -  \\ 
			
			PSR~J1614-2230  \cite{Demorest:2010bx} & -  & $1.97 \pm 0.04 $& -  \\ 
			PSR~J0537-6910 \cite{Ho:2015vza,Ho:2020vxt} & -  & $1.83 \pm 0.04 $& -  \\
				PSR~J0030+0451 \cite{Miller:2019cac} & - & $1.44 ^{+0.15}_{-0.14} $& $13.02^{+1.24}_{-1.06}$  \\ 
			GW190425 \cite{LIGOScientific:2020aai} & -  & $1.40 - 1.90 $& $11.00 - 15.00$   \\
			GW170817 \cite{LIGOScientific:2018cki, Lim:2019som,Malik:2018zcf} & -  & $1.20 - 1.70 $& $8.80 - 13.50$   \\
			\hline
		\end{tabular}
	\end{center}
\end{table}

\begin{figure}
	\centering
	\subfigure[]{\label{fig:6a}
		\includegraphics*[width=.48\linewidth, height=0.23\paperheight]{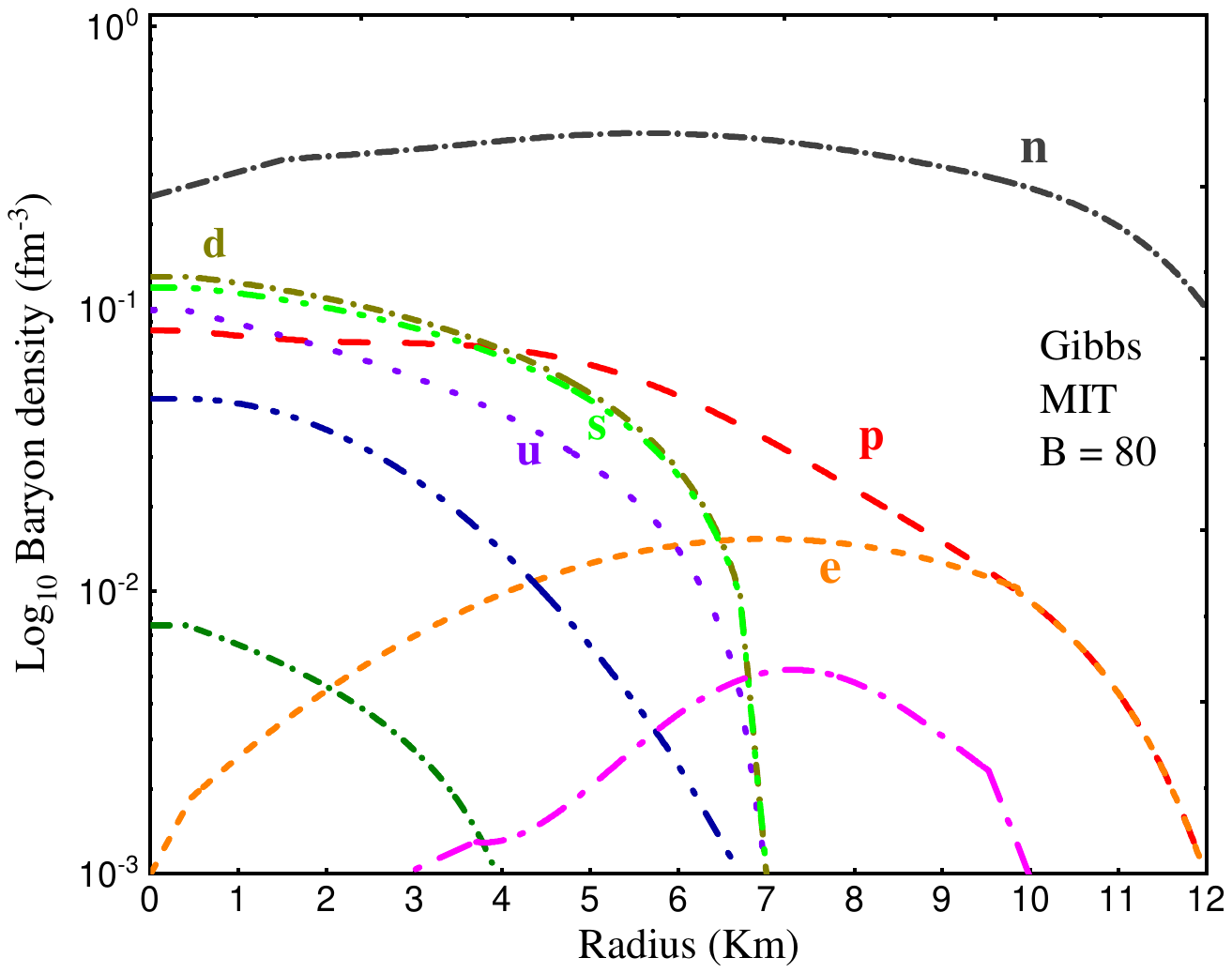}}
	\hspace{1mm}
	\subfigure[ ]{\label{fig:6b}
		\includegraphics*[width=.45\linewidth]{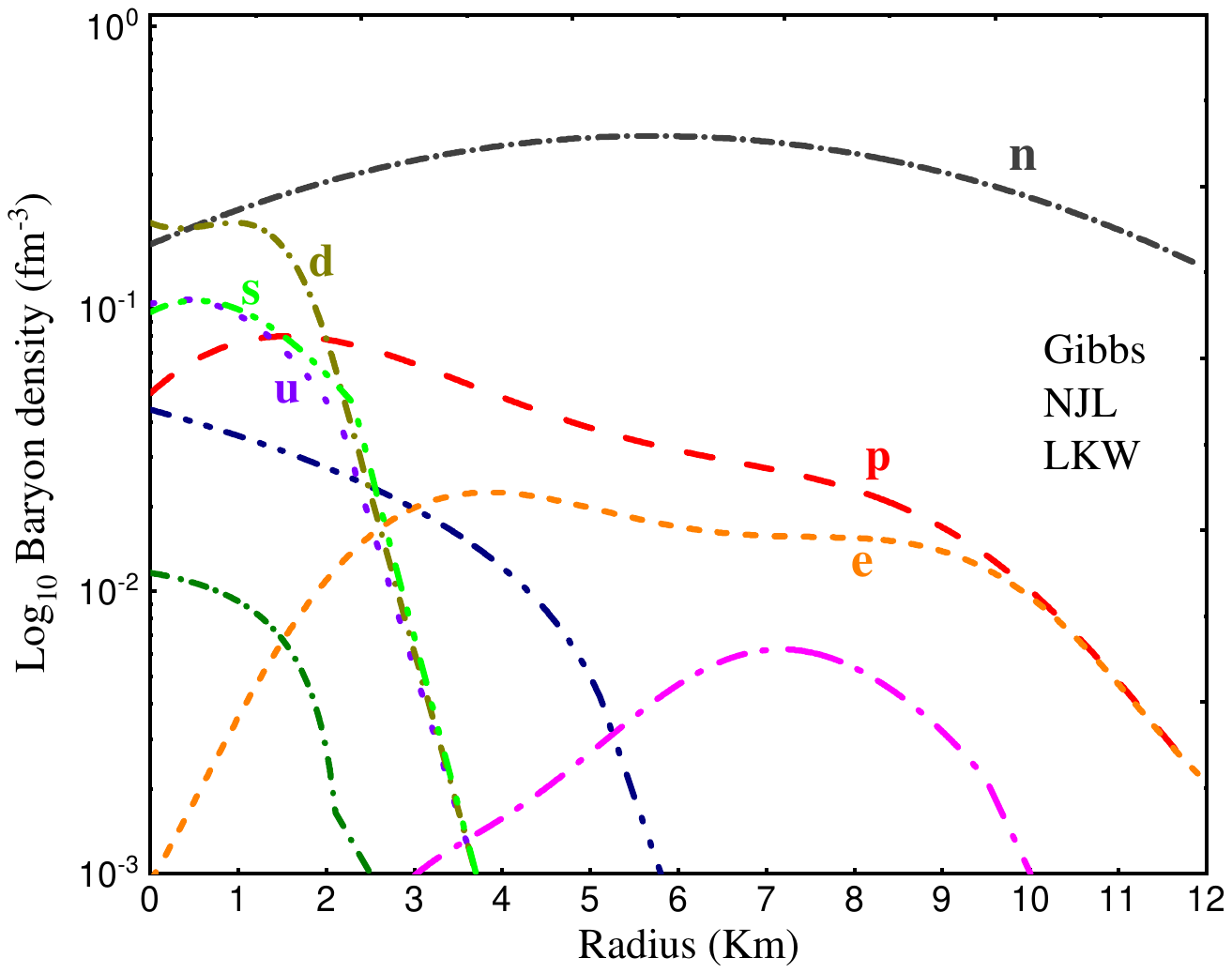}}
	
	\hspace{1mm}
	\subfigure[ ]{\label{fig:6c}
		\includegraphics*[width=.45\linewidth]{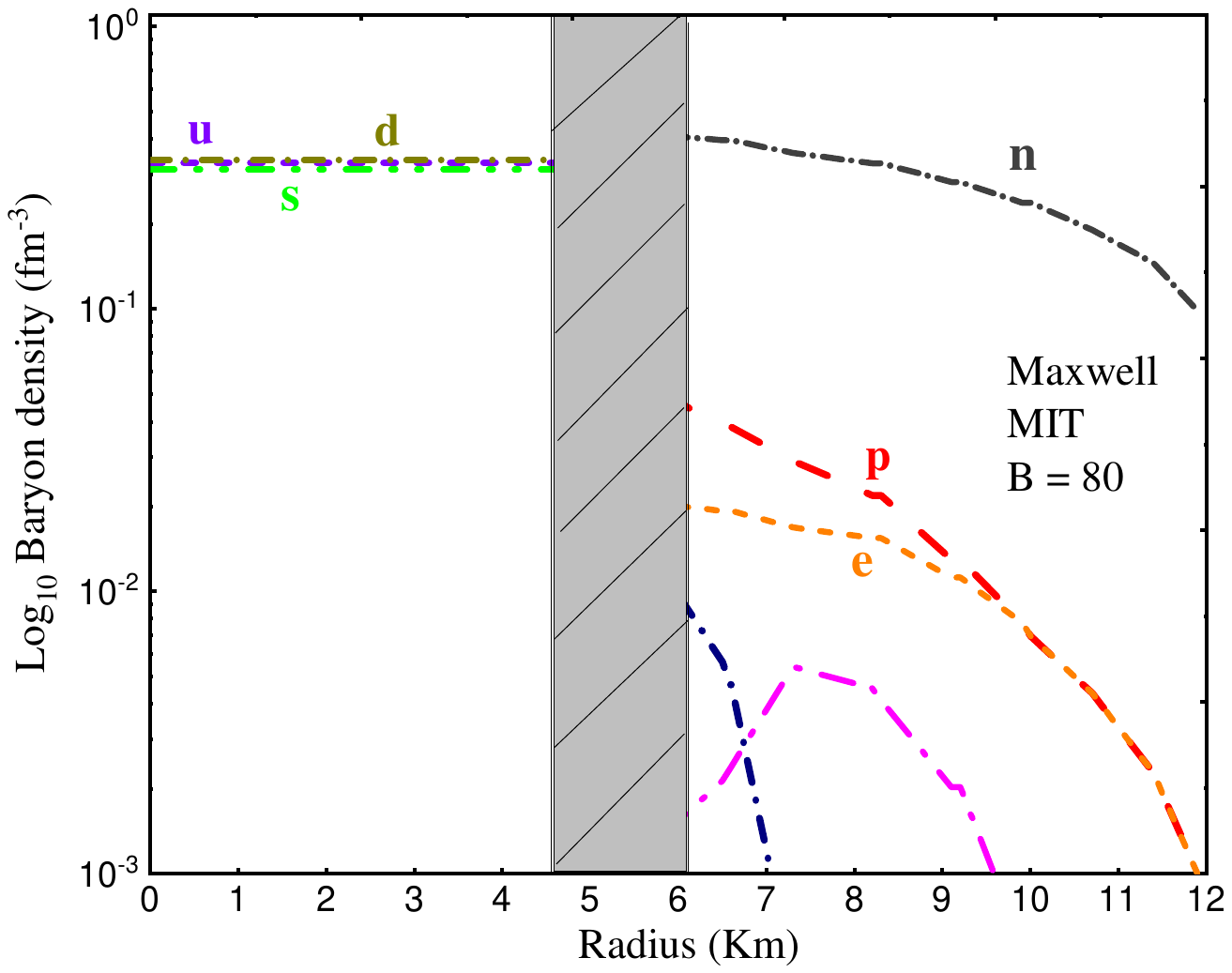}}
	
	\caption{Stellar matter for hybrid matter within the
		Gibbs and Maxwell phase transition combined with the MIT bag model and NJL model.  (a): the Gibbs construction combined with the MIT bag model for $80~\mathrm{MeV\, fm^{-3}}$, (b): the Gibbs construction with the NJL (LKW) model, and (c): the Maxwell construction with the MIT model for $80~\mathrm{MeV\, fm^{-3}}$.}
	\label{stellarmatter}
\end{figure}

The central densities listed in Table~\ref{table:m-r} further highlight structural differences between the two constructions. Maxwell configurations generally require higher central densities to approach their maximum mass, reflecting the abrupt density jump at the phase interface. Conversely, in the Gibbs construction the high-density region is distributed over an extended mixed phase, leading to smoother composition profiles and enhanced stability. An important aspect revealed by the present analysis is the impact of hyperons on the maximum mass of hybrid stars. The appearance of hyperons such as $\Lambda$ and $\Xi^-$ is also evident in both Gibbs and Maxwell solutions. Despite the softening normally associated with hyperon formation, several hybrid EOSs—including all Gibbs constructions—still produce stars exceeding 2 solar mass. This demonstrates that the mixed-phase region can compensate for hyperon softening, offering a viable path toward resolving the hyperon puzzle, showing that hyperons do not necessarily prevent the formation of massive neutron stars within the considered models.
Fig.~\ref{stellarmatter} provides complementary insight into the internal composition of the maximum-mass stars. In the Gibbs scenario, particle fractions evolve continuously with radius due to global charge neutrality and the presence of a mixed phase. Quark degrees of freedom coexist with hadronic components over a finite radial interval, and the relative appearance of hyperons and quarks depends sensitively on the model parameters. In certain cases, like Fig. \ref{fig:6a}, quark components may emerge before specific hyperonic species within the mixed phase. By contrast, the Maxwell construction produces a sharp interface separating a pure quark core from the hadronic envelope. The discontinuous behavior of particle fractions at the sharp transition radius reflects local charge neutrality and constant pressure across the boundary. This abrupt structural change is directly correlated with the reduced maximum mass and earlier gravitational instability observed in Maxwell scenarios. Taken together, these results suggest that hybrid stars with an extended mixed phase—described by the Gibbs prescription and moderate bag constants—provide the most successful description of massive neutron stars, consistently reproducing the observational requirement of  $M_{\rm max} > 2\,M_{\odot}$.

To examine the physical admissibility of the hybrid EOSs at high densities, we performed additional checks of the sound speed, causality, mechanical stability, and consistency with the perturbative-QCD (pQCD) constraint. The squared speed of sound is obtained from $c_s^2=\frac{dP}{d\varepsilon} $, where the derivative is evaluated numerically from the EOS. For the Maxwell construction the pressure remains constant across the first-order phase transition. Therefore, $c_s^2$ vanishes within the Maxwell plateau. This behavior is a characteristic feature of the first-order transition and does not imply a mechanical instability. The mechanical stability of the EOS is checked from the condition $\frac{dP}{dn_B}\geq 0$, with the equality occurring only within the Maxwell plateau.
\begin{table}[ht]
	\centering
	\caption{High-density consistency checks for the hybrid EOS.}
	\medskip
	\label{tab:consistency}
	\begin{tabular}{|lccc|}
		\hline
		
		&Gibbs + MIT  &  &  \\
		\hline
		\hline
		Quantity & $ B=60$ & $B=80$ & $B=100$ \\
		& (MeV fm$^{-3}$) & (MeV fm$^{-3}$) & (MeV fm$^{-3}$) \\
		\hline
		$n_{\rm mix}^{\rm start}$ (fm$^{-3}$) & 0.37 & 0.39 & 0.41 \\
		$n_{\rm mix}^{\rm end}$ (fm$^{-3}$)   & 1.2 & 1.2 & 1.3 \\
		$\max(c_s^2)$                         & 0.47 & 0.45 & 0.54 \\
		Causality ($c_s^2<1$)                 & Yes & Yes & Yes \\
		Mechanical stability ($\frac{dP}{dn_B}\geq 0$)                  & Yes & Yes & Yes \\
		Compatible with pQCD constraint       & Yes & Yes & Yes \\
		\hline
		\hline
		&Gibbs + NJL  &  &  \\
		\hline
		\hline
		& $ RKH$ & $HK$ & $LKW$ \\
		
		\hline
		$n_{\rm mix}^{\rm start}$ (fm$^{-3}$) & 0.54 & 0.6 & 0.55 \\
		$n_{\rm mix}^{\rm end}$ (fm$^{-3}$)   & 1.15 & 1.05 & 1.1 \\
		$\max(c_s^2)$                         & 0.54 & 0.54 & 0.54 \\
		Causality ($c_s^2<1$)                 & Yes & Yes & Yes \\
		Mechanical stability ($\frac{dP}{dn_B}\geq 0$)               & Yes & Yes & Yes \\
		Compatible with pQCD constraint       & Yes & Yes & Yes \\
		\hline
		\hline
		&Maxwell + MIT  &  &  \\
		\hline
		\hline
		Quantity & $ B=60$ & $B=80$ & $B=100$ \\
		& (MeV fm$^{-3}$) & (MeV fm$^{-3}$) & (MeV fm$^{-3}$) \\
		\hline
		$n_{\rm tr}^{\rm start}$ (fm$^{-3}$)
		& 0.40 & 0.46 & 0.52 \\
		
		$n_{\rm tr}^{\rm end}$ (fm$^{-3}$)
		& 0.54 & 0.62 & 0.69 \\

		$\max(c_s^2)$
		& 0.56 & 0.54& 0.54 \\
		
		$c_s^2$ in transition
		& 0 & 0 & 0 \\
		
		Causality ($c_s^2 < 1$)
		& Yes & Yes & Yes \\
		
		Mechanical stability ($\frac{dP}{dn_B}\geq 0$)
		& Yes & Yes & Yes \\
		
		Compatible with pQCD constraint       & Yes & Yes & Yes \\
		
		\hline
		
	\end{tabular}
\end{table}
The results of these checks are summarized in Table~\ref{tab:consistency}. For all the considered EOSs, the squared speed of sound remains below the causal limit, and no mechanical instability is found outside the phase transition region. The Gibbs and Maxwell constructions are also checked separately, taking into account the different behavior of the pressure in the mixed phase. In the Maxwell case, the vanishing sound speed within the constant pressure region is treated as a consequence of phase coexistence rather than as an instability. In addition, the high-density behavior of the hybrid EOSs is compared with the pQCD constraint obtained by Ref.~\cite{Komoltsev:2021jzg}. The comparison was performed at two representative baryon number densities, $n_B=0.176~\mathrm{fm}^{-3}$ and $n_B=1.6~\mathrm{fm}^{-3}$. At the lower reference density, the pQCD constraint yields $\mu_B=978~\mathrm{MeV}$ and $P=3.542~\mathrm{MeV~fm}^{-3}$. At $n_B=1.6~\mathrm{fm}^{-3}$, the allowed ranges are $\mu_B=1149.16$--$2195.93~\mathrm{MeV}$ and $P=36.30$--$1411.83~\mathrm{MeV~fm}^{-3}$. We evaluated the baryon chemical potential and pressure of all nine EOSs at these two densities and compared the resulting values with the corresponding pQCD constraints. At $n_B=0.176~\mathrm{fm}^{-3}$, the calculated values of $(\mu_B,P)$ are $(985, 4.8)$, for all nine EOSs, where $\mu_B$ is expressed in $MeV$ and $P$ in $\mathrm{MeV~fm}^{-3}$. At $n_B=1.6~\mathrm{fm}^{-3}$, the corresponding values are $(1598, 570.5)$, $(1598, 550.8)$, $(1598, 531.0)$, for Gibbs and maxwell construction plus MIT bag model with B = 60, 80, 100 and  $(1539, 392.1)$, $(1539, 337.3)$, $(1548, 375.1)$ for Gibbs construction plus NJL parametrization, RKH, LKW, and  HK, respectively. The results confirm that all nine EOSs satisfy the adopted pQCD consistency criteria at both reference densities, supporting their compatibility with the pQCD-constrained region in the low- and high-density regimes. Together, these checks indicate that the hybrid EOSs used in this work satisfy the required causality and stability conditions and remain physically admissible up to the central densities reached by the maximum-mass configurations.

We have also calculated the tidal deformability and Love number ($\mathcal{K}_{2}$) of neutron stars within the framework of general relativity. Gravitational waves generated during the inspiral of compact binaries-especially neutron star systems encode information about the star's tidal deformability response to the companions gravitational field. The dimensionless Love numbers, originally introduced in  Newtonian theory \cite{Love:1912} and later reformulated for relativistic systems, quantify this tidal response.  The tidal deformability parameter, denoted by $\lambda$, depends on the equation of state (EOS) through the neutron star’s radius and the Love number $\mathcal{K}_{2}$. Flanagan and Hinderer \cite{Flanagan:2007ix} derived the following expression for $\lambda$:
\begin{equation}
	\label{lambda2}
	\lambda = \frac{2}{3}  \mathcal{K}_{2}  R^{5}, 
\end{equation}
The dimensionless tidal deformability, denoted by $\varLambda$, is associated with the compactness parameter $C = \dfrac{M}{R}$ through the following relationship:
\begin{equation}
	\label{Lambdabig}
	\varLambda = \dfrac{2~\mathcal{K}_{2}}{3~ C^{5}},
\end{equation}
Let  M  and  R  denote the mass and radius, respectively, of the isolated spherical star. Throughout this analysis, we adopt natural units in which the speed of light c  and the gravitational constant G  are set to unity. The determination of the deformability parameter $\lambda$ necessitates the computation of the Love number $\mathcal{K}_{2}$, which serves as a fundamental measure of the star's deformation induced by the gravitational interaction within a binary system. The evaluation of the Love number requires the simultaneous iterative solution of the Tolman–Oppenheimer–Volkoff (TOV) equations together with the following first-order differential equation, as detailed in \cite{Hinderer:2007mb}.

The tidal deformability was evaluated at $1.4\,M_{\odot}$ for all configurations. 
At this mass, the central density remains below the onset density of the phase transition 
for all considered cases. For example, for the Gibbs construction with $B=80~\mathrm{MeV\,fm^{-3}}$, 
the transition starts at $\rho_{\mathrm{onset}}\simeq 0.39~\mathrm{fm}^{-3}$, whereas the central density 
of a $1.4\,M_{\odot}$ star is $\rho_C\simeq 0.33~\mathrm{fm}^{-3}$. Consequently, the $1.4\,M_{\odot}$ 
configurations do not contain quark matter, and all nine models probe the same underlying hadronic 
equation of state at this mass. This naturally leads to essentially identical values of the Love 
number and tidal deformability, for which we obtain $k_2\simeq 0.14$ and 
$\Lambda_{1.4}\simeq 800$. Thus, the value of $\Lambda_{1.4}$ in the present calculation should be 
regarded as a consistency check of the underlying hadronic equation of state rather than as a 
discriminator among the different quark-matter models or phase-transition constructions. In 
particular, since the $1.4\,M_{\odot}$ configurations remain entirely within the hadronic phase, 
their tidal deformability does not provide direct information on the properties of the quark 
matter sector or on the location of the phase transition. The value obtained is also comparable 
to predictions of relatively stiff hadronic equations of state reported in the literature 
\cite{Kumar:2016dks, Sedaghat:2024bnj}.

\section {CONCLUSION}\label{sec:four}
In this work, we have performed a systematic investigation of hybrid neutron stars 
by employing QCD sum rule inspired coupling constants within both Gibbs and Maxwell constructions. The analysis consistently connects the microscopic hadronic sector, single-particle potentials, equation of state behavior, particle population, and macroscopic stellar observables within a unified framework. We find that, despite the appearance of hyperons in the dense core, several Gibbs 
configurations and one Maxwell case are able to support neutron stars with masses exceeding two solar masses. This indicates that the considered coupling scheme provides sufficient stiffness to counterbalance the softening typically associated with hyperon formation, offering a viable scenario toward a partial resolution of the hyperon puzzle. 
The tidal deformability was also examined for a $1.4\,M_{\odot}$ star, yielding 
$\Lambda_{1.4}\simeq 800$ and $k_2\simeq 0.14$. Since the central density of such a star 
($\rho_C\simeq 0.33~\mathrm{fm}^{-3}$) remains below the onset of the phase transition in 
all considered configurations, this result is determined entirely by the common hadronic 
equation of state. Therefore, the identical value of $\Lambda_{1.4}$ obtained for the nine 
configurations should be interpreted as a consistency check of the hadronic sector and not 
as a constraint that discriminates among the different quark-matter models or phase-transition 
constructions. The resulting value is comparable with predictions obtained for relatively stiff 
hadronic equations of state. Compared with the hybrid-star calculations of
Ref.~\cite{Karimi:2022vxw}, the modified couplings lead to quantitative differences in the
stellar properties, including the onset of the phase transition, the maximum
stellar mass, the corresponding radius, and the internal composition. The
present analysis therefore provides a systematic assessment of the impact of
QCDSR coupling constants within the hybrid-star framework. Overall, the present study demonstrates that QCD-motivated coupling choices, when implemented consistently in hybrid star modeling, can simultaneously 
satisfy nuclear matter constraints, reproduce massive neutron stars, and yield tidal deformabilities characteristic of the underlying hadronic equation of state. These results 
highlight the importance of the underlying interaction scheme in determining 
the macroscopic behavior of compact stars and provide a coherent framework 
for further investigations under future multimessenger constraints.

	\section*{ACKNOWLEDGEMENTS}
K. Azizi thanks Iran national science foundation (INSF) for the partial financial support provided under the elites Grant No. 40405095.


\end{document}